\documentclass[10pt,twocolumn,a4paper]{IEEEtran}

\usepackage[active]{srcltx} 
\usepackage[cmex10]{amsmath}
\usepackage{amsfonts}
\usepackage{mathrsfs}
\usepackage[final]{graphics}
\usepackage[final]{graphicx}
\usepackage{amsbsy}
\usepackage{amsmath}
\usepackage{amsbsy}
\usepackage{amssymb}
\usepackage{url}
\usepackage{enumerate}
\usepackage{cite}
\usepackage{psfrag}
\usepackage{color}
\usepackage{euscript}
\usepackage[utf8]{inputenc}
\usepackage{algorithmic}
\usepackage[ruled,vlined]{algorithm2e}
\usepackage{bbm}
\usepackage{booktabs}

\newcommand{\bm}[1]{{\mathbf{#1}}}
\newcommand{\eqdef}{\triangleq}
\newcommand{\Es}{{\mathbb{E}}}          
\newcommand{\herm}{\text{H}}
\newcommand{\trasp}{\text{T}}
\newcommand{\diag}{{\text{diag}}}
\newcommand{\trace}{{\text{trace}}}
\renewcommand{\vec}{{\text{vec}}}
\newcommand{\Ml}{M_{\text{B}}}
\newcommand{\Mtilde}{\widetilde{M}}

\newcommand{\Nutilde}{\widetilde{N}_{\text{U}}}
\newcommand{\Mj}{M_{\text{J}}}
\newcommand{\Mtx}{M_{\text{TX}}}
\newcommand{\Nu}{N_{\text{U}}}
\newcommand{\rhox}{\rho_{\text{TX}}}
\newcommand{\rhobs}{\rho_{\text{B}}}
\newcommand{\rhoj}{\rho_{\text{J}}}
\renewcommand{\d}{\bm d}
\newcommand{\zerobd}{\bm 0}
\newcommand{\w}{\bm w}
\newcommand{\z}{\bm z}
\renewcommand{\c}{\bm c}
\newcommand{\Jmath}{\bm J}
\newcommand{\y}{\bm y}
\newcommand{\s}{\bm s}
\newcommand{\G}{\bm G}
\renewcommand{\r}{\boldsymbol \eta}
\newcommand{\xibold}{\boldsymbol \xi}
\newcommand{\ub}{\bm u}
\newcommand{\vb}{\bm v}
\newcommand{\ubtilde}{\widetilde{\bm u}}
\newcommand{\vbtilde}{\widetilde{\bm v}}
\newcommand{\gbtilde}{\widetilde{\bm g}}
\newcommand{\gb}{{\bm g}}
\newcommand{\htilde}{\widetilde{h}}
\renewcommand{\b}{\bm b}
\newcommand{\ax}{\bm a_{\text{TX}}}

\newcommand{\phix}{\phi_{\text{TX}}}

\newcommand{\thetax}{\theta_{\text{TX}}}
\newcommand{\varthetax}{{\vartheta}_{\text{TX}}}
\newcommand{\varphix}{{\varphi}_{\text{TX}}}

\newcommand{\taux}{\tau_{\text{TX}}}
\newcommand{\nux}{\nu_{\text{TX}}}
\newcommand{\chix}{\chi_{\text{TX}}}

\newcommand{\Lcp}{L_{\text{cp}}}
\newcommand{\Lx}{L_{\text{TX}}}
\newcommand{\Lxtilde}{\widetilde{L}_{\text{TX}}}

\newcommand{\Tx}{\text{TX}}
\newcommand{\Bs}{\text{B}}
\newcommand{\J}{\text{J}}
\newcommand{\Ue}{\text{U}}
\newcommand{\Wi}{{\bm W}^{(i)}}
\newcommand{\Ri}{{\bm R}^{(i)}}
\newcommand{\Widft}{{\bm W}_F}
\newcommand{\Tcp}{\bm T_{\text{cp}}}
\newcommand{\I}{\bm{I}}
\newcommand{\Zero}{\bm{O}}
\newcommand{\Cset}{\mathbb{C}}
\newcommand{\Rset}{\mathbb{R}}
\newcommand{\Zset}{\mathbb{Z}}
\newcommand{\pot}{\EuScript{P}}

\def\bdm#1\edm{\begin{displaymath}#1\end{displaymath}}
\def\be#1\ee{\begin{equation}#1\end{equation}}
\def\barr#1\earr{\begin{align}#1\end{align}}

\newcommand{\IeeeTIT}{{\em IEEE Trans.\ Inf. Theory\/}}
\newcommand{\IeeeTSP}{{\em IEEE Trans.\ Signal Process.\/}}
\newcommand{\IeeeTCOMM}{{\em IEEE Trans.\ Commun.\/}}
\newcommand{\IeeeCOMMLETT}{{\em IEEE Commun.\ Lett.\/}}

\newcommand{\IeeeTWC}{{\em IEEE Trans.\ Wireless Commun.\/}}
\newcommand{\IeeeJSAC}{{\em IEEE J.\ Select.\ Areas Commun.\/}}
\newcommand{\IeeeTVT}{{\em IEEE Trans.\ Veh. Technol.\/}}

\newcommand{\IeeeJSTSP}{{\em IEEE J.\ Select.\ Topics Signal Process.\/}}

\newcommand{\IeeeCOMMMAG}{{\em IEEE Commun.\ Magazine\/}}
\newcommand{\IeeeTIFS}{{\em IEEE Trans. Inf. Foren. Sec.\/}}

\begin{document}
\title{Anti-jamming beam alignment in millimeter-wave MIMO systems}

\author{Donatella~Darsena,~\IEEEmembership{Senior Member,~IEEE}
   and  Francesco~Verde,~\IEEEmembership{Senior Member,~IEEE}

\thanks{
Manuscript received September 28, 2021; 
revised April 11, 2022; 
accepted June 8, 2022.
The associate editor coordinating the review of this paper and
approving  it for publication  was Prof. N.~Gonzalez-Prelcic.
}
\thanks{
D.~Darsena is with the Department of Engineering,
Parthenope University, Naples I-80143, Italy (e-mail: darsena@uniparthenope.it).
}
\thanks{
F.~Verde is with the Department of Electrical Engineering and Information Technology, University Federico II, Naples I-80125,
Italy (e-mail: f.verde@unina.it).
}
\thanks{
The authors are also with
National Inter-University Consortium for Telecommunications (CNIT).}
}
\markboth{IEEE Transactions on Communications, Vol.~70,
No.~xx,~yy~2022}{Darsena \MakeLowercase{\textit{et al.}}:
Anti-jamming beam alignment in millimeter-wave MIMO systems}

\IEEEpubid{0000--0000/00\$00.00~\copyright~2022 IEEE}

\maketitle

\begin{abstract}
In millimeter-wave (MMW) multiple-input multiple-output (MIMO) 
communications, users and their corresponding base station (BS)
have to align their beam during both initial access and data 
transmissions to compensate for the high propa\-gation loss.
The beam alignment (BA) procedure specified for 
5th Generation (5G) New Radio (NR) has been 
designed to be fast and precise in the presence 
of non-malicious interference and noise. 
A smart jammer might exploit this weakness and may launch
an attack during the BA phase in order to  
degrade the accuracy of beam selection
and, thus, adversely impacting the end-to-end performance
and quality-of-service experienced by the users. 
In this paper, we study the effects of a jamming
attack at MMW frequencies during the  
BA procedure used to perform initial access for idle users and
adaptation/recovery for connected users. 
We show that the BA procedure adopted 
in 5G NR is extremely vulnerable to 
a smart jamming attack and, consequently, we propose 
a countermeasure based on the idea of randomized probing, 
which consists of randomly corrupting the  
probing sequence  transmitted by the BS in order
to reject the jamming signal at the UE via 
a subspace-based technique based on orthogonal projections
and jamming cancellation. 
Numerical results corroborate 
our theoretical findings  and show the very satisfactory accuracy 
of the proposed anti-jamming approach.

\end{abstract}


\section{Introduction}
\label{sec:intro}

\IEEEPARstart{T}{he} evolution of wireless radio-frequency (RF) communications has been basically
driven by the unremitting pursuit  of  large portions of unexplored
spectrum to boost the available data rates  as much as possible.
Unlike Long Term Evolution (LTE) systems, which 
mainly work below 3 GHz, 
5th Generation (5G) New Radio (NR) are allowed to
also operate in the {\em millimeter-wave (MMW)} band, 
with operating frequency from $24250$ MHz to $52600$ 
MHz \cite{3GPP-2019-6,3GPP-2019-12}.
However, MMW communications are {\em power-limited}, because of higher path losses and
blockage phenomena \cite{Rap.2013}, which demand a significant technical 
breakthrough over the LTE system.
Beamforming techniques are the standard way 
to provide the
necessary signal-to-noise ratio (SNR) gain and 
provide spatial multiplexing, by using
highly-directional beams \cite{Pau-book,Wiley}, especially in local coverage scenarios.
Directional links are realized by
antenna arrays with a large number of elements,
which are feasible at MMW signaling since,  
due to the small wavelength, it is possible to package
a large number of antenna elements at both the
base station (BS) side and the user equipment (UE) side,
implementing a massive multiple-input multiple-output (MIMO) system.
All the lower layer functions are designed in NR on the basis of 
a {\em beam-centric} philosophy: in particular, unlike
LTE, not only the user-plane channels, but also the control-plane channels are 
beamformed.

A problem arising in directional communications is how to establish, track and possibly reconfigure
beams as the UE moves, or even when the UE device is simply rotated. 
Due to mobility and blockage, the current beam pair between the BS and UE may be
blocked, resulting in a {\em beam failure event}. Beam failure could lead to {\em radio link failure
(RLF)} already defined in LTE, which is managed by a costly higher-layer reconnection
procedure. To
deal with this issue, a new set of procedures, collectively referred to as {\em beam alignment (BA)}
techniques, have been introduced in NR specifications, aimed at supporting possible
fast beam reconfiguration and tracking, preferably working at the layers $1-2$
of the protocol stack. 
The beam-centric design is a groundbreaking difference between LTE and NR, which 
makes BA strategic for control and performance
of MMW networks.
Hence, with a widespread adoption of 5G NR,
it is not difficult to imagine that BA will
become the target of all kinds of potential threats or attacks.

\IEEEpubidadjcol

\subsection{Deficiency of existing beam alignment procedure}
\label{sec:beam-management}

The task of beam management is to acquire and maintain a reliable beam pair,
i.e., a transmit angle-of-departure (AoD)
and a corresponding receive angle-of-arrival (AoA)
that jointly provides the best radio
connectivity. 
The beam management procedures specified by 
the 3rd Generation Partnership Project (3GPP) in \cite{3GPP-2018-6} and
the subsequent works \cite{Nitsche.2014,Wang.2009,Hur.2013,
Kok.2017, Song.2018, Giordani.2019,Myers.2019, Hus.2019,
Zhang.2019, Hanly.2019, Sure.2019, Li.2019, Mor.2020,
Liu.2020, Gupta.2020, Cha.2020, Ech.2021, Wang.2021,
Zhang.2021} are designed to be 
resilient to beam failure events due to mobility and blockage {\em only}. 
A noticible exception is represented by \cite{Zhang-2.2021}, where
the beam training duration, training power, and 
data transmission power are optimized to maximize 
the throughput between two legitimate nodes, while  
ensuring a covertness
constraint at a third-part node that attempts
to detect the existence of the communication. 

One serious threat to MMW network is the {\em jamming
attack during the BA phase}, for which a jammer may transmit high-power RF signals 
to induce a beam failure event 
and, thus, a RLF that prevents either idle users from accessing
the network or connected users from reconfiguration 
and tracking.  
Jamming attacks might dramatically increase the occurrence frequency
of RLFs, thus lowering the quality of service of users and 
increasing costs for system management.    
To the best of our knowledge, the jamming attack
specifically targeting at MMW links has not been considered yet
and the synthesis of effective anti-jamming schemes is an open problem. 

\subsection{Contribution and organization}

Although other solutions are possible \cite{3GPP-2018-6, Giordani.2019}, 
we focus in this paper on {\em mobile-controlled BA (MCBA)} \cite{Song.2018}
to perform initial access for idle users and
adaptation/recovery for connected users, which can be summarized as follows:

\begin{itemize}

\item 
{\em Beam sweeping}:
while all UEs stay in listening mode, 
the BS actively probes the channel by periodically broadcasting a  
beamforming codebook and a probing sequence
over reserved beacon slots in the downlink.

\item 
{\em Beam measurement}:
the UE measures the quality of the received
beamformed signals by using the received power or more sophisticated metrics,
such as the SNR.

\item 
{\em Beam determination}: the UE locally and independently identify the
best beam.

\item 
{\em Beam reporting}: the UE reports information regarding 
the best beam for successive data/control transmission or possible beam refinement
over a control uplink channel.

\end{itemize}
MCBA is highly scalable and its overhead and complexity
do not grow with the number of active users in the system.
In this paper, 
due to the rapid development 
of software-defined radio techniques, 
we explicitly account for the presence of a
{\em smart} jammer that is able to mimic
the BS signal, which is formed by
the transmit beamforming codebook and probing symbols.

Our study includes the main peculiar features 
of MMW networks. Specifically, 
according to NR physical-layer specifications, we consider  
orthogonal frequency-division multiplexing (OFDM) with cyclic prefix (CP)
as a modulation format. Moreover, we exploit the fact that,  
at MMW frequencies, propagation in dense-urban 
non-line-of-sight (NLOS) environments is only
based on a few scattering clusters, with relatively little delay/angle spreading within
each cluster \cite{Ark.2014}. In this case, the MMW channel tends to exhibit a
sparse structure in both angle and delay domains, which can be conveniently exploited
to obtain anti-jamming alignment solutions.
Finally,  due to implementation/cost constraints of fully-digital architectures, we 
rely on a realistic MMW transmit implementation \cite{Molisch.2017}, according to which 
the number of RF chains is
strictly smaller than the number of antennas.
Within such a framework, our contributions  
are the following ones:

\begin{enumerate}[(i)]

\item
We develop a detailed model of a smart jamming attack 
in MMW networks, which represents 
an important, yet open research problem.
The novelty of the proposed modeling approach rests mainly on the application
to BA of jamming transmission techniques, which
previous works attribute to communications
based on centimetre-wave (CMW) communications \cite{Lu.2014}.

\item
Existing beam sweeping procedures 
\cite{3GPP-2018-6, Nitsche.2014,Wang.2009,Hur.2013,
Kok.2017, Song.2018, Giordani.2019,Myers.2019, Hus.2019,
Zhang.2019, Hanly.2019, Sure.2019, Li.2019, Mor.2020,
Liu.2020, Gupta.2020, Cha.2020, Ech.2021, Wang.2021,
Zhang.2021} rely on 
a publicly known protocol where 
the probing symbols are known
to all UEs.
As a countermeasure against the jamming attack, 
we propose the idea of {\em randomized probing},
which consists of superimposing a random sequence on the 
known probing symbols transmitted by the BS during the beam-sweeping phase.
Such a random sequence is unknown to both the user and the jammer, 
but its subspace properties can be exploited at the UE
to reject the contribution of the jamming signal.
The idea of randomly corrupting the input data during the transmission
has been used in other contexts with different aims, such as, 
to decentralize the transmission of a space time code from a set of distributed relays \cite{Sirkeci.2007},
to boost the performance or efficiency of neural networks \cite{Zhang.2016}, 
or to overcome  the problem of pilot spoofing in CMW cellular systems \cite{Tugnait.2018}. 

\item
Performance of the proposed anti-jamming methods are validated using a
number of system parameters. It is demonstrated that 
a smart jamming attack leads to frequent beam failure events 
if no adequate countermeasures are taken. On the other hand,  
exploitation at the UE of randomized probing  
avoids beam misalignment,  
in such a way that costly 
beam recovery procedures 
are avoided while using lower-layer signaling.

\end{enumerate}

The paper is organized as follows.  
The system model of the 
BA phase under a jamming attack is described 
in Section~\ref{sec:system}.
The study of the adverse effects of the jamming attack 
on a conventional BA algorithm is reported in 
Section~\ref{sec:conv}.
The proposed countermeasure based on 
randomized probing is developed
in Section~\ref{sec:random-probing}.
Numerical results are reported 
in Section~\ref{sec:simul}.
Finally, conclusions are drawn 
in Section~\ref{sec:concl}.

\begin{figure*}[t]
\centering
\includegraphics[width=\linewidth]{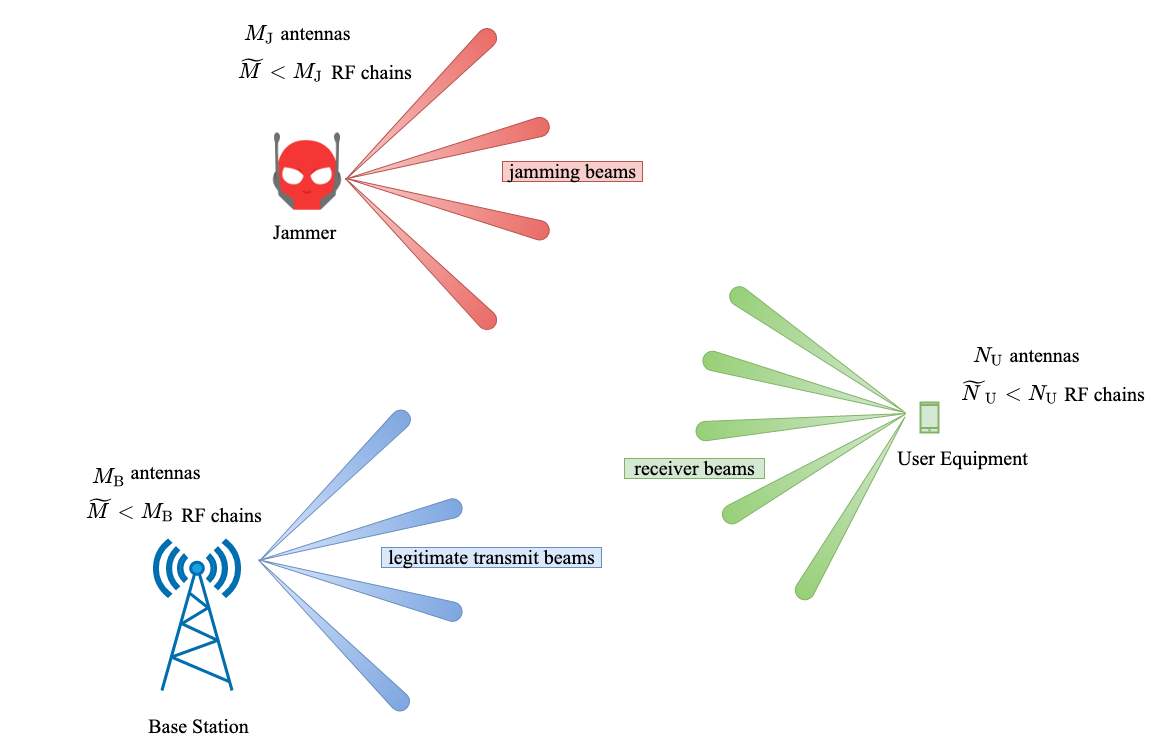}
\caption{The jamming attack on the BA procedure in MMW MIMO systems.}
\label{fig:fig_1}
\end{figure*}

\subsection{Notations}
\label{sec:pre}
Upper- and lower-case bold letters denote matrices and vectors;
the superscripts
$*$, $\trasp$, $\herm$, and $-1$
denote the conjugate,
the transpose, the Hermitian (conjugate transpose),
and the inverse of a matrix;
$\mathbb{C}$, $\mathbb{R}$, $\mathbb{Z}$, and $\mathbb{N}$ are
the fields of complex, real, integer, and natural numbers;
$\mathbb{C}^{n}$ $[\mathbb{R}^{n}]$ denotes the
vector-space of all $n$-column vectors with complex
[real] coordinates;
similarly, $\mathbb{C}^{n \times m}$ $[\mathbb{R}^{n \times m}]$
denotes the vector-space of all the $n \times m$ matrices with
complex [real] elements;
$\delta(\tau)$ is the Dirac delta;
$\delta_n$ is the Kronecker delta, i.e., $\delta_n=1$ when $n=0$
and zero otherwise;
$\jmath \eqdef \sqrt{-1}$ denotes the imaginary unit;
$\text{max}(x,y)$  returns the maximum between $x \in \Rset$ and $y \in \Rset$;
$\lceil x \rceil$
rounds $x \in \Rset$ 
to the nearest integer greater than or equal to $x$;
the (linear) convolution operator is denoted with $\ast$;
$\otimes$ stands for the Kronecker product;
$|\mathcal{A}|$ is the cardinality of the set $\mathcal{A}$;
$\mathbf{0}_{n}$, $\Zero_{n \times m}$ and $\I_{n}$
denote the $n$-column zero vector, the $n \times m$
zero matrix and the $n \times n$ identity matrix;
$\bm x \geq \zerobd_n$ [$\bm x >\zerobd_n$] denotes a vector 
$\bm x \in \Rset^n$ with non-negative [positive] entries;
$\bm{W}_n \in \mathbb{C}^{n \times n}$ is the 
unitary symmetric $n$-point inverse discrete Fourier transform 
(IDFT) matrix, whose $(m+1,p+1)$-th entry is given by
$\frac{1}{\sqrt{n}} \, e^{\jmath \frac{2 \pi}{n} m p}$ for
$m,p \in \{0,1,\ldots, n-1\}$,  
and its 
inverse $\bm{W}_n^{-1}=\bm{W}_n^\herm$ is the $n$-point 
discrete Fourier transform (DFT) matrix;
$\{\mathbf{a}\}_{\ell}$ is the $\ell$-th entry 
of $\bm{a} \in \Cset^{n}$,
for $\ell \in \{1,2,\ldots,n\}$;
$\{\mathbf{A}\}_{\ell_1,\ell_2}$ is the $(\ell_1,\ell_2)$-th entry 
of $\bm{A} \in \Cset^{n \times m}$,
for $\ell_1 \in \{1,2,\ldots,n\}$ and $\ell_2 \in \{1,2,\ldots,m\}$;
matrix $\mathbf{A} = \diag (a_{0}, a_{1}, \ldots,
a_{n-1}) \in \Cset^{n \times n}$ is diagonal;
vector $\mathbf{a} = \vec (\mathbf{A}) \in \Cset^{n \cdot m}$
is obtained by vertically stacking the columns of $\mathbf{A} \in \Cset^{m \times n}$;
let $p \ge 1$ be a real number, the $p$-norm 
(also called $\ell_p$-norm) 
of vector $\bm x \in \Cset^n$ is defined 
as $\|\bm x\|_p \eqdef \left(\sum_{i=1}^n |\{\bm x\}_i|^p\right)^{1/p}$;
$\bm{1}_{\mathcal{A}} \in \mathbb{R}^{n}$ denotes a vector whose $i$-th entry is equal to one if $i$ is contained in the set $\mathcal{A} \subseteq \{1,2,\ldots,n\}$,
otherwise is zero; $\bm{1}_n \in \mathbb{R}^{n}$ is the all-ones vector;
the support of $\bm x \in \Rset^{n}$
is the set of its nonzero entries, i.e., 
$\text{supp}(\bm x) \eqdef \big\{ i \in \{1,2,\ldots, n\} : \{\bm x\}_n \neq 0\big\}$; 
the operator $\mathscr{F}[x(n)]=\sum_{n \in \mathbb{Z}} x(n) \, e^{-\jmath \, 2 \pi \nu n}$ ($\nu \in \mathbb{R}$)
returns the Fourier transform of $x(n)$;
$\Es[\cdot]$ denotes ensemble averaging.

\section{Beam alignment model under jamming attack}
\label{sec:system}

With reference to Fig.~\ref{fig:fig_1}, we consider a MMW system employing OFDM signaling, 
with $F$ subcarriers and a CP of length $\Lcp$, which encompasses
a legitimate BS (referred to as node B), equipped with $\Ml$ antennas and $\Mtilde \ll \Ml$ RF chains, 
a generic user (referred to as node U), with $\Nu$ 
antennas and $\Nutilde \ll \Nu$ RF chains, and a jammer (referred to as node J) equipped with $\Mj$ antennas and 
$\Mtilde \ll \Mj$ RF chains. 
We study the worst case in which the jammer is perfectly aware of the BA protocol and tries 
to almost perfectly replicate the legitimate communication between the BS and 
the UE, with the scope to hinder
their corresponding beam matching, by sending {\em smart} jammer signals. 
Such an attack is hard to be detected using network monitoring tools,
since legitimate traffic on the medium will be sensed in this case \cite{Pele.2011}.
All the notations are  defined in Section~\ref{sec:pre} and the main symbols 
are summarized in Table~\ref{tab:notation}. 

\begin{table*}
\centering
\begin{tabular}{ll} 
\bf{Symbol} & \bf{Meaning}\\
 \hline
 \hline
$F$ & number of subcarriers \\
$\Lcp$ & CP length \\
$P$ & OFDM symbol length \\
$\Tx$ & subscript indicating the legitimate BS when $\Tx \equiv \Bs$  
or the jammer  when $\Tx \equiv \J$ \\
$M_{\text{TX}}$ & number of antennas at the transmitter TX\\
$N_{\text{U}}$ & number of antennas at the UE \\
$\widetilde{M}$ & number of RF chains at the BS and jammer \\
$\widetilde{N}_{\text{U}}$ & number of RF chains at the UE \\
$T$ & OFDM symbol length related to the sampling period $T_\text{c}= T/P$ \\
$W$ & BA phase length (in OFDM symbols) \\
$\mathcal{F}_i $ & set gathering the $F_i$ subcarriers assigned to the $i$-th stream \\
$d^{(k_{i,\ell})}_{\text{TX}}(s)$  & probing symbol of the $i$-th  stream transmitted 
in the $s$-th symbol interval on subcarrier $k_{i,\ell}$ \\
$\bm u^{(i,s)}_{\text{TX}}$ & transmit beamforming vector used for the $i$-th stream in the $s$-th symbol interval \\
${\bm H}_{\Tx}(\tau)$ &  impulse MIMO physical channel response between the transmitter TX and the UE \\
$\bm v^{(j,s)}$ & receive beamforming vector of the $j$-th RF chain in the $s$-th symbol interval\\
$\overline{\bm C}_\Tx^{(k)}$ & frequency-domain MIMO physical channel on subcarrier $k$ \\
$\widetilde{\bm C}^{(k)}_{\Tx}$ & frequency-domain MIMO virtual channel on subcarrier $k$ \\
$U_\Tx$ & cardinality of the angular support set probed by the transmitter TX \\
$V$ & cardinality of the angular support set sensed by the UE \\
$Q$ & number of beacon slots \\
$\gb^{(j,i)}_{\Tx}(\widetilde{s})$ & scaled version of the 
combined TX-UE beamforming vector $\gbtilde^{(j,i)}_{\Tx}(\widetilde{s})$
during the $\widetilde{s}$ beacon slot \\
$\xibold_\Tx$ & vector collecting all the unknown second-order moments of the TX-to-UE 
virtual channel\\
$\G_\Tx$ & matrix collecting all the vectors $\gb^{(j,i)}_{\Tx}(\widetilde{s})$, for
$i \in \{1,2,\ldots, \widetilde{M}\}$, $j \in \{1,2,\ldots, \widetilde{N}_\text{U}\}$,
and $\widehat{s} \in \{0,1,\ldots, Q-1\}$\\
\hline
\hline
\end{tabular}
\caption{List of the main symbols used throughout the paper.}
\label{tab:notation}
\end{table*}

\subsection{Transmit signal model}
For the sake of simplicity,  we assume that
the BS and the jammer share the same number $\Mtilde$ of RF chains and, thus, they can
transmit up to $\Mtilde$ different {\em probing} streams. 
The extension to the case in which  the jammer and the BS 
have a different number of RF chains is straightforward.
Let $\mathcal{F}_i \eqdef \{k_{i,0}, k_{i,1}, \ldots, k_{i,F_i-1}\}$ denote the 
set gathering the $F_i$ subcarriers assigned to the $i$-th stream, with 
$i \in \{1,2,\ldots,\Mtilde\}$ and 
$\sum_{i=1}^{\Mtilde} F_i \le F$. 
Such sets  $\mathcal{F}_1, \mathcal{F}_2, \ldots, \mathcal{F}_{\Mtilde}$ are disjoint,
i.e., $\mathcal{F}_{i_1} \cap \mathcal{F}_{i_2} =\emptyset$, for $i_1 \neq i_2$.
We denote with
\be
\d_{\text{TX}}^{(i)}(s) \eqdef [d^{(k_{i,0})}_{\text{TX}}(s), 
d^{(k_{i,1})}_{\text{TX}}(s), \ldots, d^{(k_{i,F_i-1})}_{\text{TX}}(s)]^\trasp \in \Cset^{F_i}
\ee
the {\em probing symbol vector} whose entry $d^{(k_{i,\ell})}_{\text{TX}}(s)$  corresponds to the $i$-th  stream, transmitted 
in the $s$-th symbol interval on subcarrier $k_{i,\ell}$, with $\Tx \in \{\Bs, \J\}$.
Since the focus is on the BA phase only, we assume that the remaining subcarriers 
$F^\text{c} \eqdef F-\sum_{i=1}^{\Mtilde} F_i$ are virtual carriers, i.e., 
subcarriers that are not used by the BS.\footnote{In practice, the BS
can use the remaining $F^\text{c}$ subcarriers to transmit 
control and data information, which is orthogonally multiplexed in frequency
with the probing symbols for BA alignment.}
After OFDM precoding, one has
\barr
\z_{\Tx}^{(i)}(s) & \eqdef [
z^{(i,0)}_{\text{TX}}(s),
z^{(i,1)}_{\text{TX}}(s),
\ldots,
z^{(i,P-1)}_{\text{TX}}(s)
]^\trasp \nonumber \\ & =
\Tcp \, \Wi \, \d_{\text{TX}}^{(i)}(s)
\nonumber
\earr
where $\Tcp \eqdef [\bm I_{\text{cp}}^{\trasp}, \bm I_F]^{\trasp} \in \Rset^{P \times F}$ accounts for CP insertion, with $\bm I_{\text{cp}} \in \Rset^{\Lcp \times F}$
collecting the last $\Lcp$ rows of $\bm I_F$, and $P \eqdef F+ \Lcp$, whereas $\Wi \in \Cset^{F \times F_i}$ represents a submatrix of the $F$-point IDFT matrix $\bm{W}_F$
(see Subsection~\ref{sec:pre}), whose elements are given by
\[
\{\Wi\}_{\ell_1+1,\ell_2+1} = \frac{1}{\sqrt{F}} \, e^{\jmath \tfrac{2 \pi}{F} \ell_1 k_{i,\ell_2}}
\]
for 
$\ell_1 \in \{0,1, \ldots, F-1\}$ and $\ell_2 \in \{0,1, \ldots, F_i-1\}$.
The vector $\z_{\Tx}^{(i)}(s)$ undergoes parallel-to-serial conversion, and the resulting sequence 
$z_{\Tx}^{(i)}(\ell)$ ($\ell \in \Zset$), defined by
$z^{(i)}_{\Tx}(s P +p) = z^{(i,p)}_{\Tx}(s)$, for $p \in \{0,1,\ldots,P-1\}$, feeds a digital-to-analog converter (DAC) having impulse response $\psi_{\text{DAC}}(t)$,
thus obtaining 
\be
x^{(i,s)}_{\Tx}(t) = \sum_{p=0}^{P-1} z^{(i,p)}_{\Tx}(s) \, \psi_{\text{DAC}}(t-sT-pT_c) 
\ee
for $t \in [s T, (s+1) T)$, with $T$ denoting the OFDM symbol length and $T_\text{c} \eqdef  T/P$.
For the BA phase, the beamforming is implemented in the analog RF domain.
We consider 
fully-connected hybrid digital
analog architecture, where each RF antenna port is connected to all
antenna elements of the array, with identity
baseband (digital) precoding matrix. 
To transmit the $i$-th probing stream, each transmitter applies an RF analog 
beamforming vector 
$\ub^{(i,s)}_{\Tx} \in \Cset^{\Mtx}$, which is assumed normalized
such that $\|\ub^{(i,s)}_{\Tx}\|_2=1$.
Hence, the baseband transmitted signal by the generic transmit terminal $\Tx$ 
during the $s$-th symbol interval is given by 
\be
\bm x_{\Tx}^{(s)}(t) =  \sum_{i=1}^{\Mtilde}  
\ub^{(i,s)}_{\Tx}  x^{(i,s)}_{\Tx}(t) \: .
\label{eq:tx-sig}
\ee
The BA phase spans a time window 
of length $W$ OFDM symbols, i.e., $W T$ seconds. 

\subsection{Physical channel model}
During the BA phase, the $\Nu \times \Mtx$ MIMO physical channel matrix between the generic transmitter  
$\Tx$ and the UE is modeled as
\be
{\bm H}_{\Tx}(\tau)  = \sum_{\ell=1}^{\Lx} \rhox(\ell) \, 
\b\left(\phix(\ell)\right) \, 
\ax^\herm\left(\thetax(\ell)\right) \, \delta\left(\tau-\taux(\ell)\right)
\label{eq:chan}
\ee
where $\Lx \ll \max\{\Mtx,\Nu\}$ denotes the number of \emph{significant} propagation 
paths,\footnote{After BA, multipath components conveying
small amount of signal power can be neglected.} 
$\rhox(\ell)$, $\taux(\ell)$,  
$\thetax(\ell) \in \left[-\tfrac{\pi}{2}, \tfrac{\pi}{2}\right)$ and 
$\phix(\ell) \in \left[-\tfrac{\pi}{2}, \tfrac{\pi}{2}\right)$  
are the channel gain, the time delay, the 
angle-of-departure (AoD), and 
angle-of-arrival (AoA) of the 
the $\ell$-th multipath component, respectively, 
whereas $\ax\left(\thetax(\ell)\right) \in \Cset^{\Mtx}$ and 
$\b\left(\phix(\ell)\right) \in \Cset^{\Nu}$ 
are the array responses of transmitter $\Tx$ and UE, respectively,
which depend on the array geometry and they are 
parameterized by the AoD and AoA, respectively.
We have invoked the customary assumption that the communication bandwidth 
of the transmitted signals is much smaller than the carrier frequency $f_0$, such that 
the array responses can be assumed independent of frequency. 
We have also assumed that, for $\ell 
\in \{1, 2, \ldots, \Lx\}$, the channel gain $\rhox(\ell)$, AoD $\thetax(\ell)$ and AoA $\phix(\ell)$ are time-invariant during 
the BA phase, i.e., over $W$ OFDM symbols, 
since they typically vary on time intervals much longer than the channel coherence time.

Since each propagation path is approximately equal to the sum of independent 
micro-scatterers contributions, having same time delay and AoA-AoD, the channel gains 
$\rhox(\ell)$, for $\ell\in \{1, 2, \ldots, \Lx\}$ and $\Tx \in \{\Bs, \J\}$, 
can thus be modeled as independent zero-mean 
complex circular Gaussian random variables (RVs)
({\em  uncorrelated Rayleigh scattering environment}), with variances $\sigma_{\Tx}^2(\ell)$,  
where 
$\rhobs(\ell)$  is statistically independent of $\rhoj(\ell)$. 

\subsection{Receive signal model}

Since  the noise in the receiver is mainly introduced
by the RF chain electronics (filter, mixer, and A/D conversion),
we neglect ambient noise external to the 
radio receiving system  \cite{ITU}.
From \eqref{eq:tx-sig} and \eqref{eq:chan}, it follows that the baseband equivalent received signal at the UE antenna array during the BA phase reads as follows
\begin{multline}
\r(t)  = \sum_{\Tx \in \{\Bs, \J\}} \sum_{h} 
\int  {\bm H}_{\Tx}(\tau) \, \bm x_{\Tx}^{(h)}(t-\tau) \, {\rm d} \tau 
\\  =  
 \sum_{\Tx \in \{\Bs, \J\}} 
\sum_{h} \sum_{\ell=1}^{\Lx} \sum_{i=1}^{\Mtilde} 
\rhox(\ell) \, g_{\Tx}^{(i,h)}(\ell)  
\\ \cdot
 x^{(i,h)}_{\Tx}\left(t-\taux(\ell)\right)
 \, \b\left(\phix(\ell)\right)
\end{multline}
where $g_{\Tx}^{(i,h)}(\ell) \eqdef \ax^\herm\left(\thetax(\ell)\right) \, \ub^{(i,h)}_{\Tx} \in \Cset$ 
denotes the {\em beamforming gain} along the
$\ell$-th propagation path, in the $h$-th symbol interval, at the transmitter 
$\Tx$ for the $i$-th RF chain.

Hereinafter, we assume perfect frequency and time synchronization.
Recalling that the UE is equipped with $\Nutilde$ RF chains, after power splitting by a factor of $\Nutilde$
and anti-aliasing filtering,  
the baseband equivalent received signal at the output of the $j$-th RF chain can be written as
\begin{multline}
y_\text{a}^{(j,s)}(t) = \frac{1}{\sqrt{\Nutilde}} \left\{\left[\vb^{(j,s)}\right]^{\herm} \r(t) \right\} \ast \psi_{\text{ADC}}(t) + 
w_\text{a}^{(j,s)}(t) 
\\ =
\frac{1}{\sqrt{\Nutilde}} \sum_{\Tx \in \{\Bs, \J\}} 
\sum_{h=s-1}^{s} \sum_{\ell=1}^{\Lx} \sum_{i=1}^{\Mtilde} 
\sum_{p=0}^{P-1} \rhox(\ell) \, f_{\Tx}^{(j,s)}(\ell) \, g_{\Tx}^{(i,h)}(\ell)  
\\ \cdot
z^{(i,p)}_{\Tx}(h) \, 
\psi_\text{a}\left(t-\taux(\ell)-hT-pT_\text{c}\right) + w_\text{a}^{(j,s)}(t) 
\label{eq:ric}
\end{multline}
for $t \in [s T, (s+1) T)$, with $s \in \{0,1,\ldots, W-1\}$, and $j \in \{1,2,\ldots,\Nutilde\}$,
where we denote with $\vb^{(j,s)} \in \Cset^{\Nu}$ the beamforming vector of the $j$th RF chain at the UE side,
normalized such that $\|\vb^{(j,s)}\|_2=1$,  $\psi_{\text{ADC}}(t)$ is the impulse response of the analog-to-digital converter (ADC), 
$f_{\Tx}^{(j,s)}(\ell) \eqdef \left[{\vb^{(j,s)}}\right]^{\herm} \b\left(\phix(\ell)\right) \in \Cset$
represents the {\em array gain} of the $j$th RF chain along the $\ell$-th propagation path at the UE side, 
$w_\text{a}^{(j,s)}(t)$ is complex circular white Gaussian noise  at the output of the $j$th RF chain, statistically independent of $\d_{\text{TX}}^{(i)}(h)$, 
for $\Tx \in \{\Bs, \J\}$, $h \in \{s-1,s\}$, and $i \in \{1,2,\ldots,\Mtilde\}$, and, finally, 
$\psi_\text{a}(t)  \eqdef \psi_{\text{DAC}}(t) \ast \psi_{\text{ADC}}(t)$
is a unit-energy Nyquist pulse-shaping filter.
We have also assumed in \eqref{eq:ric} that 
$L_\psi \, T_\text{c}+ \taux(\ell) < T$, for each
$\ell \in \{1,2,\ldots, \Lx\}$ and $\Tx \in \{\Bs, \J\}$,
with $L_\psi$ being the duration of $\psi_\text{a}(t)$ (in sampling periods),
so as the signal $y_\text{a}^{(j,s)}(t)$ is impaired only by the interblock
interference (IBI) of the symbol transmitted in the previous 
signaling interval $t \in [(s-1) T, s T)$ and, thus, the integer $h$ is restricted to
the binary set $\{s-1,s\}$.

The continuous-time signal \eqref{eq:ric} is sampled with rate $1/{T_c}$ at time instants $t_{s,q} \eqdef s T + q T_\text{c}$, for $q \in \{0,1,\ldots,P-1\}$. 
Let $y^{(j,q)}(s) \eqdef y_\text{a}^{(j,s)}(t_{s,q})$ be the discrete-time counterpart of \eqref{eq:ric}, one gets
\begin{multline}
y^{(j,q)}(s) = \sum_{\Tx \in \{\Bs, \J\}} \sum_{h=s-1}^{s} \sum_{i=1}^{\Mtilde} \sum_{p=0}^{P-1} 
z^{(i,p)}_{\Tx}(h)   \\ \cdot  c_{\Tx}^{(j,i,s,h)}\left((s-h)P +q-p\right) + w^{(j,q)}(s)
\label{eq:sig}
\end{multline}
with
\begin{multline}
c_{\Tx}^{(j,i,s,h)}(r) \eqdef \frac{1}{\sqrt{\Nutilde}} \sum_{\ell=1}^{\Lx} 
\rhox(\ell) \, f_{\Tx}^{(j,s)}(\ell) \\  \cdot g_{\Tx}^{(i,h)}(\ell) \, \psi_\Tx\left(r-\nux(\ell),\ell\right)
\end{multline}
where $r \in \Zset$, we have defined $\psi_\Tx(r,\ell) \eqdef \psi_\text{a}\left(r T_c-\chix(\ell)\right)$ and  
$\taux(\ell)= \nux(\ell) \, T_\text{c} + \chix(\ell)$, 
with integer delay $\nux(\ell)$
and fractional delay $\chix(\ell) \in [0, T_\text{c})$, 
and noise sample $w^{(j,q)}(s) \eqdef w_\text{a}^{(j,s)}(t_{s,q})$.
Under the assumption that the CP is sufficiently long, i.e., 
$\Lcp \ge L_\psi+(\tau_\text{max}-\tau_\text{min})/T_\text{c}$,
where $\tau_\text{min} \eqdef \min_{\Tx,\ell} \taux(\ell)$ and 
$\tau_\text{max} \eqdef \max_{\Tx,\ell} \taux(\ell)$, 
the IBI contributions in \eqref{eq:sig}, which are represented by the 
addends corresponding to $h=s-1$, can be suppressed through CP removal.
Therefore, by removing the
first $\Lcp$ samples (corresponding to the CP), gathering the obtained data into 
the vector $\bm y^{(j)}(s) \eqdef [y^{(j,\Lcp)}(s), y^{(j,\Lcp+1)}(s), \ldots, y^{(j,P-1)}(s)]^\trasp \in \Cset^F$, and 
accounting for \eqref{eq:sig}, one obtains, for $j \in \{1,2,\ldots,\Nutilde\}$, the following IBI-free vector model
\be
\bm y^{(j)}(s) = \sum_{\Tx \in \{\Bs, \J\}} \sum_{i=1}^{\Mtilde} \text{circ}\left( \bm c_{\Tx}^{(j,i)}(s)\right) \, \Wi \, \d_{\text{TX}}^{(i)}(s)
+ \w^{(j)}(s)
\ee
where $\text{circ}\left( \bm c_{\Tx}^{(j,i)}(s)\right) \in \Cset^{F \times F}$ is the circulant \cite{Horn} channel matrix 
having 
\begin{multline}
\bm c_{\Tx}^{(j,i)}(s) \eqdef [c_{\Tx}^{(j,i,s,s)}(\Lcp), c_{\Tx}^{(j,i,s,s)}(\Lcp-1), \\ \ldots,
c_{\Tx}^{(j,i,s,s)}(0), 0, \ldots, 0] \in \Cset^{1 \times F}
\nonumber
\end{multline}
as its first row, whereas the additive noise is given by 
$\bm w^{(j)}(s) \eqdef [w^{(j,\Lcp)}(s), w^{(j,\Lcp+1)}(s), \ldots, w^{(j,P-1)}(s)]^\trasp \in \Cset^F$.
At this point, performing the DFT 
of $\bm y^{(j)}(s)$ and recalling that circulant matrices are diagonalized by the DFT \cite{Horn}, the frequency-domain received data vector assumes the form
\be
\overline{\y}^{(j)}(s) = \sum_{\Tx \in \{\Bs, \J\}} \sum_{i=1}^{\Mtilde} \diag\left(
\overline{\c}_\Tx^{(j,i)}(s)\right) \, \Ri \, 
\d_{\text{TX}}^{(i)}(s) + \overline{\w}^{(j)}(s)
\label{eq:sig-y}
\ee
for $j \in \{1,2,\ldots,\Nutilde\}$, where $\overline{\y}^{(j)}(s) \eqdef  \Widft^\herm \, \bm y^{(j)}(s) \in \Cset^F$, 
$\Widft^\herm$ is the $F$-point 
DFT matrix (see Subsection~\ref{sec:pre}), the vector $\overline{\c}_\Tx^{(j,i)}(s) \eqdef [\overline{c}_\Tx^{(j,i,0)}(s), \overline{c}_\Tx^{(j,i,1)}(s), \ldots, \overline{c}_\Tx^{(j,i,F-1)}(s)]^{\trasp} \in \Cset^{F}$
gathers the frequency-domain channel samples given by
\begin{multline}
\overline{c}_\Tx^{(j,i,k)}(s) =
\frac{1}{\sqrt{\Nutilde}} \sum_{\ell=1}^{\Lx} 
\rhox(\ell) \, f_{\Tx}^{(j,s)}(\ell) \, g_{\Tx}^{(i,s)}(\ell) \\ \cdot 
e^{-\jmath \frac{2 \pi}{F} k \, \nux(\ell)} \, \Psi_\Tx\left(\frac{k}{F},\ell \right)
\label{eq:c}
\end{multline}
for $k \in \{0,1,\ldots, F-1\}$,
with $\Psi_{\Tx}(\nu,\ell ) = \mathscr{F}[\psi_{\Tx}(r,\ell)]$
being the Fourier transform of $\psi_{\Tx}(r,\ell)$
with respect to $r$, whereas $\Ri \eqdef \bm{W}_F^\herm \, \Wi \in \Rset^{F \times F_i}$ is a binary ($0/1$) matrix 
that extends the probing vector $\d_{\text{TX}}^{(i)}(s)$ 
with the insertion of $F-F_i$ zeros 
on the subcarriers belonging to the set $\mathcal{F}_{i}^\text{c}$, 
which is the complement of $\mathcal{F}_{i}$ with respect to the subcarrier set 
$\{0,1,\ldots, F-1\}$, and, finally, 
$\overline{\w}^{(j)}(s) \eqdef \Widft^\herm \, \w^{(j)}(s)$ is modeled as zero-mean complex circular Gaussian  
with  $\Es[\overline{\w}^{(j_1)}(s_1) \, \overline{\w}^{(j_2)}(s_2)]= 
\sigma_w^2 \, \delta_{j_1-j_2} \, \delta_{s_1-s_2} \, \I_{F}$.

By substituting into \eqref{eq:c} the beamforming gain 
$g_{\Tx}^{(i,s)}(\ell) = \ax^\herm\left(\thetax(\ell)\right) \, \ub^{(i,s)}_{\Tx}$
and array gain $f_{\Tx}^{(j,s)}(\ell) = \left[{\vb^{(j,s)}}\right]^{\herm} \b\left(\phix(\ell)\right)$, 
one has the compact expression
$\overline{c}_\Tx^{(j,i,k)}(s) = \left[{\vb^{(j,s)}}\right]^{\herm} \overline{\bm C}_\Tx^{(k)} \, \ub^{(i,s)}_{\Tx}$,  
where matrix $\overline{\bm C}_\Tx^{(k)} \in \Cset^{\Nu \times \Mtx}$ is given by
\begin{multline}
\overline{\bm C}_\Tx^{(k)} \eqdef \frac{1}{\sqrt{\Nutilde}} 
\sum_{\ell=1}^{\Lx} \rhox(\ell) \, \Psi_\Tx\left(\frac{k}{F},\ell \right)  \\
\cdot e^{-\jmath \frac{2 \pi}{F} k \, \nux(\ell)} \, \b\left(\phix(\ell)\right) \, \ax^\herm\left(\thetax(\ell)\right) 
 \label{eq:phy}
 \end{multline}
for $k \in \{0,1,\ldots, F-1\}$. Consequently, after some straightforward manipulations, eq. \eqref{eq:sig-y} admits the 
equivalent form:
\begin{multline}
\overline{\y}^{(j)}(s) = \sum_{\Tx \in \{\Bs, \J\}} 
\sum_{i=1}^{\Mtilde}  \left[\I_F \otimes {\vb^{(j,s)}}\right]^{\herm}  
\overline{\bm C}_\Tx \left[\I_F \otimes \ub^{(i,s)}_{\Tx}\right] 
\\ \cdot \Ri \, 
\d_{\text{TX}}^{(i)}(s) + \overline{\w}^{(j)}(s) 
\label{eq:sig-y-new}
\end{multline}
for $j \in \{1,2,\ldots,\Nutilde\}$, where we have defined the block-diagonal matrix
\be
\overline{\bm C}_\Tx \eqdef \diag\left(\overline{\bm C}_\Tx^{(0)}, \overline{\bm C}_\Tx^{(1)},
\ldots, \overline{\bm C}_\Tx^{(F-1)}\right) \in \Cset^{(\Nu F) \times (\Mtx F)} \: .
\ee

\subsection{Virtual channel model}
The highly directional nature of propagation, together with the large number of antennas employed in MMW systems,
makes {\em virtual}
or {\em canonical model} of MIMO channel \cite{Sayeed, Heath, Song.2018} a natural choice
for our framework.
Specifically, we assume that the BS, jammer and receiver are equipped with a \emph{uniform linear array (ULA)} and 
we assume that the UE is in the far-field of both the transmitters. In this case, let $d_{\Tx}$ and $d_{\Ue}$ denote the antenna spacing at the generic transmit node $\Tx$ and the UE,
respectively, the (normalized) array vectors are given by 
\begin{multline}
\bm a_{\Tx}\left(\thetax(\ell)\right)  \equiv \widetilde{\bm a}_{\Tx}\left(\varthetax(\ell)\right) 
\eqdef \frac{1}{\sqrt{\Mtx}} \left[1, e^{-\jmath \, 2 \pi {\varthetax(\ell)}}, 
\right. \\ \left. e^{-\jmath \, 4 \pi {\varthetax(\ell)}},  \ldots, e^{-\jmath \, 2 \pi {\varthetax(\ell)}(\Mtx-1)}\right]^{\trasp} 
\\
\bm b\left(\phix(\ell)\right)  \equiv \widetilde{\bm b}\left(\varphix(\ell)\right) \eqdef  \frac{1}{\sqrt{\Nu}} \left[1, e^{-\jmath \, 2 \pi {\varphix(\ell)}}, 
\right. \\ \left. e^{-\jmath \, 4 \pi {\varphix(\ell)}}, \ldots, e^{-\jmath \, 2 \pi {\varphix(\ell)}(\Nu-1)}\right]^{\trasp} 
\end{multline}
where the {\em normalized spatial angles} ${\varthetax}(\ell)$ and ${\varphix}(\ell)$ are related to the physical 
AoD $\thetax(\ell)$ and physical AoA $\phix(\ell)$ 
through the relations ${\varthetax}(\ell) \eqdef (d_{\Tx}/{\lambda_0}) \, \sin \thetax(\ell)$ and 
${\varphix}(\ell) \eqdef (d_{\Ue}/{\lambda_0}) \, \sin \phix(\ell)$, respectively,
whereas the wavelength is $\lambda_0=c/f_0$, with  
$c$ being the speed of the light in the medium.
Hereinafter, we set 
 $d_{\Tx} =d_{\Ue}=\lambda_0/2$ for simplicity, which implies that 
$|{\varthetax}(\ell)| \le 1/2$ and $|{\varphix}(\ell)| \le 1/2$, 
for any $\ell \in \{1,2,\ldots, \Lx\}$.
 
The virtual representation of the physical channel 
can be obtained \cite{Sayeed-pro} by uniformly  
sampling \eqref{eq:phy} in the AoD-AoA-delay $3$-D domain at the Nyquist rate 
$(\Delta \varthetax, \Delta \varphix, \Delta \nu_{\Tx})=(1/{\Mtx}, 1/{\Nu}, T_\text{c})$,
where $1/T_\text{c}$ is (approximately) the two-sided bandwidth of the OFDM signal. 
Therefore, the virtual representation of the channel matrix \eqref{eq:phy} is 
approximately given by\footnote{The effect 
of the frequency-domain coefficient $\Psi_\Tx\left({k}/{F},\ell \right) $ disappears 
in the sampled representation \eqref{eq:C-mat} of the physical model  \eqref{eq:phy} if
the pulse $\psi_\text{a}(t)$ satisfies the Nyquist criterion.}
\begin{multline}
\overline{\bm C}_\Tx^{(k)} = \sum_{n=0}^{\Nu-1} \sum_{m=0}^{\Mtx-1}  
\sum_{\widetilde{\ell}=0}^{\Lxtilde-1} 
\widetilde{C}_{\Tx}^{(n,m,\widetilde{\ell})} \, \widetilde{\bm b}\left(\frac{n}{\Nu}-\frac{1}{2}\right) \\ \cdot
\widetilde{\bm a}_{\Tx}^{\herm}\left(\frac{m}{\Mtx}-\frac{1}{2}\right) \, e^{-\jmath \, \frac{2 \pi}{F} 
k \, \widetilde{\ell}\, T_\text{c}}
\label{eq:C-mat}
\end{multline}
for $k \in \{0,1,\ldots, F-1\}$, where the $\Nu \, \Mtx \, \Lxtilde$
{\em virtual channel coefficients} $\{\widetilde{C}_{\Tx}^{(n,m,\widetilde{\ell})}\}$
completely characterize the channel matrix \eqref{eq:phy},
with $\Nu$, $\Mtx$, and $\Lxtilde \eqdef \lceil \nu_{\Tx,\text{max}}/T_\text{c} \rceil+1$
denoting the maximum number of \emph{resolvable} AoAs, AoDs, and delays in 
the AoD-AoA-delay $3$-D domain, and, finally,  
$\nu_{\Tx,\text{max}} \eqdef \max_{\ell} \nu_{\Tx}(\ell)$.
It is worth noting that each virtual coefficient $\widetilde{C}_{\Tx}^{(n,m,\widetilde{\ell})}$ is approximately equal to the sum of the 
complex gains of all the physical paths whose angles and delays belong to the resolution bin of dimension
$\Delta \varthetax \times \Delta \varphix \times \Delta \nu_{\Tx}$
centered around the sampling point $(m/{\Mtx}-1/2, n/{\Nu}-1/2, \widetilde{\ell} \, T_\text{c})$ in the 
AoD-AoA-delay $3$-D domain.
We assume that each $\widetilde{C}_{\Tx}^{(n,m,\widetilde{\ell})}$
is a circularly symmetric complex Gaussian random
variable. According to the central limit theorem, this is a
reasonable assumption if there is a sufficiently large number
of unresolvable physical paths contributing to each 
$\widetilde{C}_{\Tx}^{(n,m,\widetilde{\ell})}$.
Moreover, 
if $\Mtx$, $\Nu$, and $1/T_\text{c}$ are sufficiently large, the virtual channel
coefficients are approximately statistically independent (see \cite{Sayeed}
for details).
Henceforth, we assume that 
the channel coefficients $\widetilde{C}_{\Bs}^{(n,m,\widetilde{\ell})}$ 
and $\widetilde{C}_\text{J}^{(n,m,\widetilde{\ell})}$ are mutually
independent zero-mean uncorrelated
RVs, i.e.,  
$\Es[\widetilde{C}_{\Tx}^{(n_1,m_1,\widetilde{\ell}_1)} \, 
\{\widetilde{C}_{\Tx}^{(n_2,m_2,\widetilde{\ell}_2)}\}^*]=
\widetilde{\sigma}_{\Tx}^2(n_1,m_1,\ell_1) \, 
\delta_{n_1-n_2} \, \delta_{m_1-m_2} \, \delta_{\widetilde{\ell}_1-\widetilde{\ell}_2}$,
for $\Tx \in \{\Bs, \J\}$, where 
$\widetilde{\sigma}_{\Tx}^2(n,m,\ell)$ is related to the variances 
of the physical channel gains via virtual partitioning of the
paths \cite{Sayeed}.

Let $\Jmath_r \eqdef \diag(1, e^{\jmath \pi}, e^{\jmath 2 \pi}, \ldots, e^{\jmath \pi (r-1)}) \in \Rset^{r}$,
it is readi\-ly verified that
$\widetilde{\bm a}_{\Tx}\left({m}/{\Mtx}-1/2\right)=\Jmath_{\Mtx}^* \, \widetilde{\bm a}_{\Tx}\left({m}/{\Mtx}\right)$ and 
$\widetilde{\bm b}\left({n}/{\Nu}-1/2\right)=\Jmath_{\Nu}^* \, 
\widetilde{\bm b}\left({n}/{\Nu}\right)$.
By observing that $\Jmath_{\Mtx}^* \, \widetilde{\bm a}_{\Tx}\left({m}/{\Mtx}\right)$ and 
$\Jmath_{\Nu}^* \, \widetilde{\bm b}\left({n}/{\Nu}\right)$ in \eqref{eq:C-mat}
turn out to be the $(m+1)$-th column and $(n+1)$-th column of the 
$\Mtx$-point DFT matrix $\bm W_{\Mtx}^{\herm}$ and 
the $\Nu$-point DFT matrix $\bm W_{\Nu}^{\herm}$ matrix, respectively, 
for $m \in \{0, 1, \ldots, \Mtx-1\}$ and $n \in \{0,1,\ldots, \Nu-1\}$, 
the channel matrix \eqref{eq:C-mat} can be expressed in a more compact form as
\be
\overline{\bm C}_\Tx^{(k)} = \Jmath_{\Nu} \, \bm W_{\Nu}^{\herm}  \, \widetilde{\bm C}^{(k)}_{\Tx} \, \bm W_{\Mtx} \, \Jmath_{\Mtx}^*
\label{eq:sparse}
\ee
for $k \in \{0,1,\ldots, F-1\}$, with 
\be
\widetilde{\bm C}^{(k)}_{\Tx} \eqdef \sum_{\widetilde{\ell}=0}^{\Lxtilde-1}
\widetilde{\bm C}^{(\widetilde{\ell})}_{\Tx} \, e^{-\jmath \, \frac{2 \pi}{F} k \, 
\widetilde{\ell} \, T_\text{c}}
\label{eq:Ctildekappa}
\ee
where the $(n+1,m+1)$-th entry of $\widetilde{\bm C}^{(\widetilde{\ell})}_{\Tx} \in \Cset^{\Nu \times \Mtx}$ is given by 
$\widetilde{C}_{\Tx}^{(n,m,\widetilde{\ell})}$, 
for $n \in \{0,1,\ldots, \Nu-1\}$ and $m \in \{0, 1, \ldots, \Mtx-1\}$.
Representation \eqref{eq:sparse} is of paramount importance 
since the {\em virtual channel matrix} 
$\widetilde{\bm C}^{(k)}_{\Tx}$ captures the {\em sparse} nature of the
MMW MIMO channel: indeed, wireless
channels with clustered multipath components tend to
have far fewer than $\Nu \, \Mtx \, \Lxtilde$
virtual channel coefficients
when operate at large bandwidths and symbol
durations and/or with massive number of antennas. 
It can be verified numerically that, 
as the number of transmit $\Mtx$ and receive $\Nu$ antennas increases,  
the matrix $\widetilde{\bm C}^{(k)}_{\Tx}$  becomes more and more sparse.

At this point, substituting \eqref{eq:sparse} into \eqref{eq:sig-y-new}
and using the mixed-product property of the Kronecker product \cite{Horn}, 
the received signal can be conveniently rewritten in terms of the virtual
channel as 
\begin{multline}
\overline{\y}^{(j)}(s) = \sum_{\Tx \in \{\Bs, \J\}} 
\sum_{i=1}^{\Mtilde}  \left[\I_F \otimes {\vbtilde^{(j,s)}}\right]^{\herm}  
\widetilde{\bm C}_\Tx \left[\I_F \otimes \ubtilde^{(i,s)}_{\Tx}\right] 
\\ \cdot \Ri \, 
\d_{\text{TX}}^{(i)}(s) + \overline{\w}^{(j)}(s)\:,  
\quad \text{for $j \in \{1,2,\ldots,\Nutilde\}$}
\label{eq:sig-y-fin}
\end{multline}
where we have defined 
$\vbtilde^{(j,s)}  \eqdef \bm \Jmath_{\Nu} \, \bm W_{\Nu} \, {\vb^{(j,s)}} \in \Cset^{\Nu}$, 
$\widetilde{\bm C}_\Tx  \eqdef \diag\left(\widetilde{\bm C}_\Tx^{(0)}, \widetilde{\bm C}_\Tx^{(1)}, \ldots, \widetilde{\bm C}_\Tx^{(F-1)}\right) \in \Cset^{(\Nu F) \times (\Mtx F)}$, 
and 
$\ubtilde^{(i,s)}_{\Tx}  \eqdef \bm \Jmath_{\Mtx} \, \bm W_{\Mtx} \, \ub^{(i,s)}_{\Tx} 
\in \Cset^{\Mtx}$.

The two sets 
\[
\left \{\ubtilde^{(i,s)}_{\text{B}}, \, \text{for $i \in \{1,2,\ldots,\Mtilde\}$
and $s \in \{0,1,\ldots, W-1\}$}\right\}
\]
and 
\[
\left\{\ubtilde^{(i,s)}_{\text{J}}, \, \text{for $i \in \{1,2,\ldots,\Mtilde\}$ and 
$s \in \{0,1,\ldots, W-1\}$}\right\}
\]
 represent the {\em transmit beamforming codebooks} of 
the BS and the jammer, respectively, which define the directions along
which the transmit beam patterns $\{\ub^{(i,s)}_{\text{B}}\}$ and 
$\{\ub^{(i,s)}_{\text{J}}\}$ send the legitimate and jamming signal power, respectively.
On the other hand, the set 
\[
\left\{\vbtilde^{(j,s)}, \text{for $j \in \{1,2,\ldots,\Nutilde\}$ and $s \in \{0,1,\ldots, W-1\}$}\right\}
\]
represents the {\em receive beamforming codebook} of the UE, which defines 
the directions
from which the receiver beam patterns $\{\vb^{(j,s)}\}$ collect the overall
signal power.

\subsection{Beacon slot model}
Recalling that the probing vectors 
$\d_{\text{TX}}^{(1)}(s), \d_{\text{TX}}^{(2)}(s), \ldots, \d_{\text{TX}}^{(\Mtilde)}(s)$ 
corresponding to the $\Mtilde$ streams 
are allocated to disjoint subcarrier sets, i.e.,  $\mathcal{F}_{i_1} \cap \mathcal{F}_{i_2} =\emptyset$ for $i_1 \neq i_2$, we focus on the $F_i$ subcarriers assigned to the $i$-th stream, 
with $i \in \{1,2,\ldots,\Mtilde\}$, by picking up only the entries 
$\overline{y}^{(j, k_{i,0})}(s), \overline{y}^{(j, k_{i,1})}(s), \ldots, \overline{y}^{(j, k_{i,F_i-1})}(s)$
of the received
vector $\overline{\y}^{(j)}(s)$ at the output of the $j$-th RF chain with indices in the 
set $\mathcal{F}_{i} = \{k_{i,0}, k_{i,1}, \ldots, k_{i,F_i-1}\}$.
On such subcarriers the contribution of the probing vectors 
$\d_{\text{TX}}^{(i')}(s)$ for $i' \neq i$ is zero.
So doing, from \eqref{eq:sig-y-fin}, the $i$-th probing signal  
received during the $s$-th data block 
on subcarrier $k_{i,\ell}$ is 
\begin{multline}
\overline{y}^{(j, k_{i,\ell})}(s)  =  \sum_{\Tx \in \{\Bs, \J\}}
[\vbtilde^{(j,s)}]^{\herm} \, 
\widetilde{\bm C}_\Tx^{(k_{i,\ell})}  \\ \cdot \ubtilde^{(i,s)}_{\Tx} \, 
d^{(k_{i,\ell})}_{\text{TX}}(s) + \overline{w}^{(j, k_{i,\ell})}(s) 
\\  = 
\sum_{\Tx \in \{\Bs, \J\}}  [\gbtilde^{(j,i,s)}_{\Tx}]^\herm  \, 
\vec\left(\widetilde{\bm C}_\Tx^{(k_{i,\ell})}\right) \\ \cdot 
d^{(k_{i,\ell})}_{\text{TX}}(s) + \overline{w}^{(j, k_{i,\ell})}(s)
\label{eq:sig-y-scal}
\end{multline}
with $j \in \{1,2,\ldots,\Nutilde\}$, $i \in \{1,2,\ldots,\Mtilde\}$, and 
$\ell \in \{0,1,\ldots, F_i-1\}$, 
where  
$\overline{w}^{(j, k_{i,0})}(s), \overline{w}^{(j, k_{i,1})}(s), \ldots, \overline{w}^{(j, k_{i,F_i-1})}(s)$
are the entries of $\overline{\w}^{(j)}(s)$ with indices in the 
set $\mathcal{F}_{i}$, 
we have used the identity  
$[\vbtilde^{(j,s)}]^{\herm} \, 
\widetilde{\bm C}_\Tx^{(k_{i,\ell})}  \, \ubtilde^{(i,s)}_{\Tx} =
\{[\ubtilde^{(i,s)}_{\Tx}]^\trasp \otimes [\vbtilde^{(j,s)}]^{\herm}\} \, 
\vec(\widetilde{\bm C}_\Tx^{(k_{i,\ell})})$ \cite{Brewer}, 
and $\gbtilde^{(j,i,s)}_{\Tx} \eqdef 
[\ubtilde^{(i,s)}_{\Tx}]^* \otimes \vbtilde^{(j,s)}
\in \Cset^{\Mtx \Nu}$ represents the combined TX-UE beamforming vector.

\begin{figure*}[t]
\centering
\includegraphics[width=\linewidth]{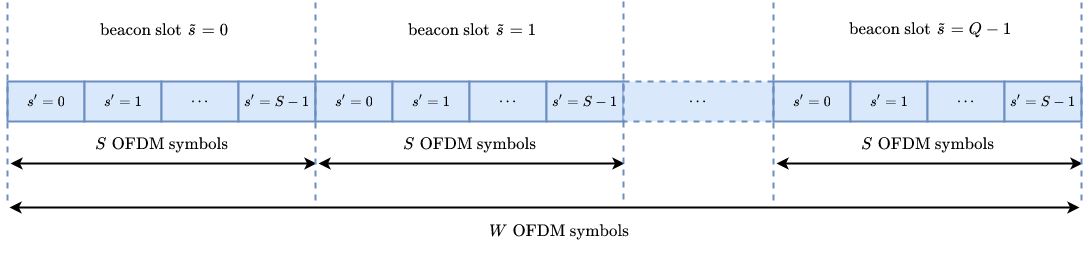}
\caption{The BA phase spans a time window 
of $W$ OFDM symbols, which is divided into $Q$ beacons slots of $S$ OFDM symbols.}
\label{fig:fig_2}
\end{figure*}

As depicted in Fig.~\ref{fig:fig_2} at the top of the next page, 
the BA phase is divided in $Q$ {\em beacon slots} of duration equal to 
$S$ consecutive OFDM blocks, i.e., $W=Q \, S$. 
At this point, let us denote with $\overline{y}^{(j, k_{i,\ell},s')}(\widetilde{s}) \eqdef 
\overline{y}^{(j, k_{i,\ell})}(\widetilde{s} \, S + s')$
the polyphase decomposition of the received data \eqref{eq:sig-y-scal}
with respect to $S$, for $\widetilde{s} \in \{0,1,\ldots, Q-1\}$ and 
$s' \in \{0, 1, \ldots, S-1\}$.
It assumed that 
the beamforming vectors $\ubtilde^{(i,s)}_{\Tx}$ and ${\vbtilde^{(j,s)}}$ are constant in each beacon slot, but they may
vary from a beacon slot to another, i.e., 
$\ubtilde^{(i,\widetilde{s} \, S + s')}_{\Tx} \equiv 
\ubtilde^{(i)}_{\Tx}(\widetilde{s})$ and 
$\vbtilde^{(j,\widetilde{s} \, S + s')} \equiv 
\vbtilde^{(j)}(\widetilde{s})$.
In this case, according to \eqref{eq:sig-y-scal}, one has 
\begin{multline}
\overline{y}^{(j, k_{i,\ell},s')}(\widetilde{s}) = \sum_{\Tx \in \{\Bs, \J\}} 
[\gbtilde^{(j,i)}_{\Tx}(\widetilde{s})]^\herm \, \vec\left(\widetilde{\bm C}_\Tx^{(k_{i,\ell})}\right) 
\\ \cdot 
d^{(k_{i,\ell},s')}_{\text{TX}}(\widetilde{s}) + \overline{w}^{(j, k_{i,\ell},s')}(\widetilde{s})
\label{eq:overline_y}
\end{multline}
where 
$\gbtilde^{(j,i)}_{\Tx}(\widetilde{s})  \eqdef 
[\ubtilde^{(i)}_{\Tx}(\widetilde{s})]^* \otimes \vbtilde^{(j)}(\widetilde{s})
\in \Cset^{\Mtx \Nu}$, 
$d^{(k_{i,\ell},s')}_{\text{TX}}(\widetilde{s})  \eqdef 
d^{(k_{i,\ell})}_{\text{TX}}(\widetilde{s} \, S + s')$, and noise sample
$ \overline{w}^{(j, k_{i,\ell},s')}(\widetilde{s})  \eqdef 
\overline{w}^{(j, k_{i,\ell})}(\widetilde{s} \, S + s')$.
We are considering the case of perfect beacon synchronization between BS and UE,
as well as between the jammer and UE. Such an assumption is reasonable in practice since 
beacon slots are periodically repeated and, thus, terminals can easily acquire perfect knowledge of 
the start epoch of each beacon slot \cite{Song.2018}.

\subsection{Structure of the beamforming codebooks}
\label{sec:codebook}

To ensure spatial coverage, 
the size of the transmit and receive beamforming codebooks 
is proportional to the number of transmit and receive antennas.
Therefore, for large-scale array in
MMW communication, exhaustive search \cite{Ak2014},
although guaranteeing to select the optimal beam,
introduces unacceptable beam training overhead.
On the other hand, hierarchical schemes \cite{Alk.2014}
require a non-trivial coordination
among the UEs and the BS, which is difficult to have at the
initial channel acquisition stage. 
Even though the proposed anti-jamming strategy can be 
applied to many available beamforming schemes, 
we resort herein to pseudo-random beamforming codebooks 
\cite{Song.2018,Kok.2017,Ven.2017}, which 
do not require interaction between the BS and each UE,
and their overhead and complexity
do not grow with the number of active users in the system.
According to these schemes, the beamforming vectors of 
the transmitter $\text{TX}$ and UE are
\be
\ubtilde^{(i)}_{\text{TX}}(\widetilde{s})  = \frac{\bm{1}_{\mathcal{U}_{\text{TX}}^{(i)}(\widetilde{s})}}{\sqrt{U_\Tx}}
\quad \text{and} \quad
\vbtilde^{(j)}(\widetilde{s}) = \frac{\bm{1}_{\mathcal{V}^{(j)}(\widetilde{s})}}
{\sqrt{V}}
\label{eq:beam}
\ee
respectively, where the {\em angular support sets} $\mathcal{U}_{\text{TX}}^{(i)}(\widetilde{s}) \subseteq \{1, 2, \ldots, \Mtx\}$ and 
$\mathcal{V}^{(j)}(\widetilde{s}) \subseteq \{1, 2, \ldots, \Nu\}$
of cardinality 
$U_\Tx \eqdef \left|\,\mathcal{U}_{\text{TX}}^{(i)}(\widetilde{s})\right|$ and 
$V \eqdef \left|\mathcal{V}^{(j)}(\widetilde{s})\right|$ 
collect the angles in the virtual
beamspace channel representation that are probed 
by $\text{TX}$ and sensed by the UE, respectively.
So doing, the combined TX-UE beamforming vector in 
\eqref{eq:overline_y} reads as
\be
\gbtilde^{(j,i)}_{\Tx}(\widetilde{s}) = \frac{\bm{1}_{\mathcal{U}_{\text{TX}}^{(i)}(\widetilde{s})} \otimes \bm{1}_{\mathcal{V}^{(j)}(\widetilde{s})}}{\sqrt{U_\Tx} \, \sqrt{V}}
\label{eq:gtildesimpl} 
\ee
whose entries are equal to $0$ or $1$ depending on 
the elements of the sets  
$\mathcal{U}_{\text{TX}}^{(i)}(\widetilde{s})$ and 
$\mathcal{V}^{(j)}(\widetilde{s})$.

The transmit beamforming codebook of the BS and the 
receive beamforming codebook of the UE are  indeed
pseudo-random since  
$\mathcal{U}_{\Bs}^{(i)}(\widetilde{s})$, 
for $i \in \{1,2,\ldots,\Mtilde\}$, 
and $\mathcal{V}^{(j)}(\widetilde{s})$, 
for $j \in \{1,2,\ldots,\Nutilde\}$, 
are generated in a pseudo-random manner, 
for each beacon slot $\widetilde{s} \in \{0,1,\ldots, Q-1\}$.
At each beacon slot, the subsets 
$\mathcal{U}_{\Bs}^{(i)}(\widetilde{s})$, for 
$i \in \{1,2,\ldots,\Mtilde\}$, are perfectly known 
at the UE, whereas 
$\mathcal{V}^{(j)}(\widetilde{s}) $, for 
$j \in \{1,2,\ldots,\Nutilde\}$, are locally  and independently 
generated by the UE.
As regards the jammer, the transmit beamforming codebook 
$\mathcal{U}_{\text{J}}^{(i)}(\widetilde{s})$  is assumed to be unknown 
at the UE, for each beacon slot
and $i \in \{1,2,\ldots,\Mtilde\}$. The impact of the choice of the jamming codebook 
on the BA procedure between the BS and the UE is discussed in Section~\ref{sec:conv}.

\subsection{Probing symbols of the BS in conventional schemes}
\label{sec:probing}

Conventional BA schemes 
\cite{3GPP-2018-6, Nitsche.2014,Wang.2009,Hur.2013,
Kok.2017, Song.2018, Giordani.2019,Myers.2019, Hus.2019,
Zhang.2019, Hanly.2019, Sure.2019, Li.2019, Mor.2020,
Liu.2020, Gupta.2020, Cha.2020, Ech.2021, Wang.2021,
Zhang.2021} do not account for jamming attacks.
In such {\em jammer-unaware} methods, the BS transmits known probing symbols 
during each beacon slot:
\be
d^{(k_{i,\ell})}_{\Bs}(s) = \sqrt{\pot_{\Bs}} \, 
t^{(k_{i,\ell})}(s) 
\label{eq:d-conv}
\ee
where $t^{(k_{i,\ell})}(s) \in \Cset$ is a publicly known symbol corresponding to the $i$-th  stream
in the $s$-th block on subcarrier $k_{i,\ell}$, 
with $|t^{(k_{i,\ell})}(s)|^2=1$,  for 
$\ell \in \{0,1,\ldots, F_i-1\}$, and $\pot_\text{B}$ 
is the available power per symbol at the BS.
In Section~\ref{sec:random-probing}, we will suitably modify
the transmission scheme \eqref{eq:d-conv} to confer 
anti-jamming capabilities to the BA procedure.

\subsection{Probing symbols of the jammer}
\label{sec:probing}
The probing symbols transmitted by the jammer are essentially a noisy version
of the publicly known probing symbols $\{t^{(k_{i,\ell})}(s) \}$ and they 
are modeled as 
\be
d^{(k_{i,\ell})}_{\text{J}}(s) = 
\sqrt{(1-\gamma_\text{J}) \, \pot_\text{J}} \, 
t^{(k_{i,\ell})}(s)
+\sqrt{\gamma_\text{J} \, \pot_\text{J}} \, 
r^{(k_{i,\ell})}_{\text{J}}(s)
\label{eq:d-j}
\ee
where $\pot_\text{J}$ is the available
power per symbol of the jammer and
each stream $\{r^{(k_{i,\ell})}_{\text{J}}(s)\}$
is modeled  as a sequence of 
zero-mean unit-variance 
independent and identically distributed (i.i.d.) complex
circular RVs.
For the sake of generality, 
we have introduced in \eqref{eq:d-j}
a power factor $0 \le \gamma_\text{J} \le 1$ that 
allows us to account for
different jamming attacks. 
In extreme cases, the jammer may exclusively transmit  
known probing symbols, i.e., $\gamma_\text{J}=0$ or,
on the other hand, it might send in the air noise only, i.e., 
$\gamma_\text{J}=1$. In the intermediate case $0 < \gamma_\text{J} <1$, 
the jammer could decide to split its available power 
between known probing symbols and intentional noise.

\section{Jammer-unaware beam alignment}
\label{sec:conv}

In this section, we show what is the impact of transmit beamforming codebook
of the jammer on the BA acquisition performance when the BS uses the conventional
probing scheme \eqref{eq:d-conv} in the presence of the jamming attack.
In this situation, the received signal by the UE is obtained
by substituting \eqref{eq:d-conv} and \eqref{eq:d-j} into \eqref{eq:overline_y}, thus obtaining
\begin{multline}
\overline{y}^{(j, k_{i,\ell},s')}(\widetilde{s}) = \sqrt{\pot_\text{B}}\, 
\htilde^{(j, i,  k_{i,\ell})}_{\Bs}(\widetilde{s}) \, t^{(k_{i,\ell}, s')}(\widetilde{s})
\\
+ 
\htilde^{(j, i,  k_{i,\ell})}_{\text{J}}(\widetilde{s}) 
\left[ \sqrt{(1-\gamma_\text{J}) \, \pot_\text{J}} \, 
t^{(k_{i,\ell})}(s)
\right. \\ \left.
+\sqrt{\gamma_\text{J} \, \pot_\text{J}} \, 
r^{(k_{i,\ell}, s')}_{\text{J}}(\widetilde{s})
\right] + \overline{w}^{(j, k_{i,\ell},s')}(\widetilde{s})
\label{eq:overline_y_mod-conv}
\end{multline}
where $\widetilde{s} \in \{0,1,\ldots, Q-1\}$ and $s' \in \{0, 1, \ldots, S-1\}$, 
with
$\htilde^{(j, i,  k_{i,\ell})}_{\Tx}(\widetilde{s}) \eqdef [\gbtilde^{(j,i)}_{\Tx}(\widetilde{s})]^\herm \, \vec\left(\widetilde{\bm C}_\Tx^{(k_{i,\ell})}\right)$, 
$t^{(k_{i,\ell},s')}(\widetilde{s}) \eqdef 
t^{(k_{i,\ell})}(\widetilde{s} \, S + s')$, 
and 
$r^{(k_{i,\ell},s')}_{\text{J}}(\widetilde{s}) \eqdef 
r^{(k_{i,\ell})}_{\text{J}}(\widetilde{s} \, S + s')$. 
With reference to the beacon slot $\widetilde{s} \in \{0,1,\ldots, Q-1\}$ (see Fig.~\ref{fig:fig_2}), 
by stacking $S$ consecutive samples \eqref{eq:overline_y_mod-conv} into the vector
\begin{multline}
\overline{\bm y}^{(j, k_{i,\ell})}(\widetilde{s}) \eqdef [\overline{y}^{(j, k_{i,\ell},0)}(\widetilde{s}),
\overline{y}^{(j, k_{i,\ell},1)}(\widetilde{s}), 
\\ \ldots, \overline{y}^{(j, k_{i,\ell},S-1)}(\widetilde{s})]^{\trasp} \in \Cset^S
\nonumber
\end{multline}
one obtains
\begin{multline}
\overline{\bm y}^{(j, k_{i,\ell})}(\widetilde{s}) = \sqrt{\pot_\text{B} } \,  
\htilde^{(j, i,  k_{i,\ell})}_{\Bs}(\widetilde{s}) \, \bm t^{(k_{i,\ell})}(\widetilde{s}) 
 \\ + \htilde^{(j, i,  k_{i,\ell})}_{\text{J}}(\widetilde{s})
\left[ \sqrt{(1-\gamma_\text{J}) \, \pot_\text{J}} \, 
\bm t^{(k_{i,\ell})}(\widetilde{s})
\right. \\ \left.
+\sqrt{\gamma_\text{J} \, \pot_\text{J}} \, 
\bm r_{\text{J}}^{(k_{i,\ell})}(\widetilde{s})
\right]
+ \overline{\bm w}^{(j, k_{i,\ell})}(\widetilde{s})
\label{eq:yvec}
\end{multline}
where 
\barr
\bm t^{(k_{i,\ell})}(\widetilde{s})   & \eqdef [t^{(k_{i,\ell},0)}(\widetilde{s}), t^{(k_{i,\ell},1)}(\widetilde{s}),  
\nonumber \\ & \hspace{20mm} \ldots, t^{(k_{i,\ell},S-1)}(\widetilde{s})]^{\trasp} \in \Cset^S
\nonumber \\
\bm r_{\text{J}}^{(k_{i,\ell})}(\widetilde{s})   & \eqdef [r_{\text{J}}^{(k_{i,\ell},0)}(\widetilde{s}), r_{\text{J}}^{(k_{i,\ell},1)}(\widetilde{s}),  
\nonumber \\ & \hspace{20mm} \ldots, r_{\text{J}}^{(k_{i,\ell},S-1)}(\widetilde{s})]^{\trasp} \in \Cset^S
\nonumber \\
\overline{\bm w}^{(j, k_{i,\ell})}(\widetilde{s})    & \eqdef [\overline{w}^{(j, k_{i,\ell},0)}(\widetilde{s}),
\overline{w}^{(j, k_{i,\ell},1)}(\widetilde{s}), \nonumber \\ & \hspace{20mm} \ldots, \overline{w}^{(j, k_{i,\ell},S-1)}(\widetilde{s})]^{\trasp} \in \Cset^S \: .
\nonumber
\earr

The strongest multipath components of the legitimate channel correspond to the entries 
with large variance of
the channel matrix $\overline{\bm C}_\Bs^{(k)}$, which is defined
in \eqref{eq:C-mat} and represented by \eqref{eq:sparse}, 
for $k \in \{0,1,\ldots, F-1\}$.
To identify the variance of such components and, thus, achieve successfully BA,  
several objective functions can be used in a jammer-unaware approach
\cite{Nitsche.2014,Wang.2009,Hur.2013,
Kok.2017, Song.2018, Giordani.2019,Myers.2019, Hus.2019,
Zhang.2019, Hanly.2019, Sure.2019, Li.2019, Mor.2020,
Liu.2020, Gupta.2020, Cha.2020, Ech.2021, Wang.2021,
Zhang.2021}.
Herein, we focus on the second-order objective function introduced 
in \cite{Song.2018} that can be expressed as 
\be
P^{(j, i)}(\widetilde{s}) = \frac{1}{S \, F_i}  
\sum_{\ell=0}^{F_i-1} 
\Es\left[\left \|\overline{\bm y}^{(j, k_{i,\ell})}(\widetilde{s}) \right \|_2^2\right] 
\label{eq:object}
\ee
which represents the {\em (normalized) mean received power} of the $i$-th data stream 
at the output of the $j$-th RF chain 
during the $\widetilde{s}$-th beacon slot,
where the expectation is also evaluated with respect to the random probing
symbols transmitted by the jammer.
By substituting \eqref{eq:yvec} into \eqref{eq:object} and invoking the statistically independence 
among channels, random sequences, and noise, one has
\be
P^{(j, i)}(\widetilde{s}) = 
[\gb^{(j,i)}_{\Bs}(\widetilde{s})]^\trasp \, \xibold_\Bs 
+ 
[\gb^{(j,i)}_{\text{J}}(\widetilde{s})]^\trasp \, \xibold_\text{J} +
\sigma^2_w
\label{eq:object-1}
\ee
where we have additionally observed that
\barr
\Es\left[ |\htilde^{(j, i,  k_{i,\ell})}_{\Tx}(\widetilde{s}) |^2 \right] & =
[\gbtilde^{(j,i)}_{\Tx}(\widetilde{s})]^\herm \,  \bm R_{\widetilde{\bm C}_\Tx} \, 
\gbtilde^{(j,i)}_{\Tx}(\widetilde{s}) 
\nonumber \\ & = 
[\gb^{(j,i)}_{\Tx}(\widetilde{s})]^\trasp \, \xibold_\Tx
\label{eq:Eh}
\earr
with 
\begin{multline}
\bm R_{\widetilde{\bm C}_\Tx} \eqdef \Es\left[\vec\left(\widetilde{\bm C}_\Tx^{(k_{i,\ell})}\right) \, \vec^\herm\left(\widetilde{\bm C}_\Tx^{(k_{i,\ell})}\right) \right]
\\ \in \Cset^{(\Mtx \Nu) \times (\Mtx \Nu)}
\nonumber 
\end{multline}
being the covariance matrix of the vectorized beamspace representation of the channel matrix.
It is worth noting that, under the assumption that the 
virtual channel coefficients are uncorrelated., the matrix
$\bm R_{\widetilde{\bm C}_\Tx}$ is diagonal with some dominant components along the 
diagonal and, according to  \eqref{eq:Ctildekappa}, 
it turns out to be independent of the subcarrier index $k_{i,\ell}$. 
In \eqref{eq:Eh}, we set $\gb^{(j,i)}_{\Tx}(\widetilde{s}) \eqdef 
\sqrt{U_\Tx} \, \sqrt{V} \, \gbtilde^{(j,i)}_{\Tx}(\widetilde{s})$, 
with the combined TX-UE beamforming vector $\gbtilde^{(j,i)}_{\Tx}(\widetilde{s})$ given by
\eqref{eq:gtildesimpl},
whereas 
\begin{multline}
\xibold_\Tx \eqdef \frac{\pot_{\Tx}}{U_\Tx \, V} \left[\{\bm R_{\widetilde{\bm C}_\Tx}\}_{1,1},  \{\bm R_{\widetilde{\bm C}_\Tx}\}_{2,2}, 
\right.  \\  \left. \ldots, 
\{\bm R_{\widetilde{\bm C}_\Tx}\}_{\Mtx \Nu, \Mtx \Nu} \right]^\trasp 
\in \Rset^{\Mtx \Nu} \: .
\label{xibold}
\end{multline}
Within this section, we assume that $\sigma_w^2$ is known at the UE for beam determination.
We remember that the vector $\gb^{(j,i)}_{\Bs}(\widetilde{s})$ is known at the UE, 
for all values of $j \in \{1,2,\ldots,\Nutilde\}$, $i \in \{1,2,\ldots,\Mtilde\}$, 
and $\widetilde{s} \in \{0,1,\ldots, Q-1\}$. On the other hand, 
$\gb^{(j,i)}_\text{J}(\widetilde{s})$ is {\em unknown} at the UE,
since it does not have knowledge of both the transmit number of antennas $\Mj$
and beamforming codebook $\mathcal{U}_{\text{J}}^{(i)}(\widetilde{s})$ of the jammer.
The unknown vector $\xibold_\Bs$ has to be estimated 
to identify the AoA-AoD directions of the strongest
scatterers regarding the BS-to-UE channel. To this aim, the UE can collect all the available power measurements
in the vector 
\begin{multline}
\bm p  \eqdef [P^{(1, 1)}(0),  \ldots, P^{(\Nutilde, \Mtilde)}(0), 
\\ P^{(1, 1)}(1),  \ldots, P^{(\Nutilde, \Mtilde)}(1), \ldots, \\ P^{(1, 1)}(Q-1),  \ldots, P^{(\Nutilde, \Mtilde)}(Q-1)]^\trasp
\\ = \G_\Bs \, \xibold_\Bs + \G_\text{J} \, \xibold_\text{J} + \sigma^2_w \, \bm{1}_{\Mtilde \Nutilde Q}
\label{eq:p}
\end{multline}
with 
\begin{multline}
\G_\Tx \eqdef [\gb^{(1,1)}_{\Tx}(0), \ldots, \gb^{(\Nutilde, \Mtilde)}_{\Tx}(0), 
\\ \gb^{(1,1)}_{\Tx}(1), \ldots, \gb^{(\Nutilde, \Mtilde)}_{\Tx}(1), \ldots, \\
\gb^{(1,1)}_{\Tx}(Q-1), \ldots, \gb^{(\Nutilde, \Mtilde)}_{\Tx}(Q-1)]^\trasp \\ \in \Rset^{(\Mtilde \Nutilde Q) \times (\Mtx \Nu)} \:.
\label{eq:Gtx}
\end{multline}
The model \eqref{eq:p} represents 
a high-dimensional system in which the number of unknowns
$M_\Bs \, \Nu$ is at least of the same order of magnitude as the number of observations 
$\Mtilde \, \Nutilde \, Q$ or, even, $M_\Bs \, \Nu \gg \Mtilde \, \Nutilde \, Q$, in which 
case one cannot hope to recover the desired vector $\xibold_\Bs$
if it does not exhibit any particular structure. However, the vector
$\xibold_\Bs$ is sparse and its entries are non-negative, i.e., 
$\xibold_\Bs \geq \zerobd_{M_\Bs \Nu}$.
If the UE is unaware of the jamming attack, an estimate of $\xibold_\Bs$ can be obtained
by solving the {\em non-negative least-squares (NNLS)} problem:
\begin{multline}
\widehat{\xibold}_\Bs=\arg \min_{\xibold_\Bs^\star \in \Rset^{M_\Bs \Nu}}
\left \|\bm p- \G_\Bs \, \xibold_\Bs^\star-\sigma^2_w \, \bm{1}_{\Mtilde \Nutilde Q}\right\|_2^2\: ,
\\ \quad \text{subject to $\xibold_\Bs^\star \ge \zerobd_{M_\Bs \Nu}$}
\label{eq:nnls}
\end{multline}
which is a convex optimization problem that can be solved efficiently \cite{Kim.2010}.
In the absence of the jamming attack, under mild conditions on the matrix $\G_\Bs$,  
the non-negativity constraint $\xibold_\Bs^\star \geq \zerobd_{M_\Bs \Nu}$ alone suffices for sparse recovery
of $\xibold_\Bs$, without the need to employ sparsity-promoting regularization terms \cite{Sla.2013}.
The minimization program \eqref{eq:nnls} is directly implemented in
MATLAB as the function \verb+lsqnonneg+, which executes the active-set
algorithm of Lawson and Hanson \cite{Bjorck}.

In practice, the NNLS problem to be solved comes from replacing $\bm p$
in \eqref{eq:nnls} with the corresponding estimate
\begin{multline}
\widehat{\bm p}  \eqdef [\widehat{P}^{(1, 1)}(0),  \ldots, \widehat{P}^{(\Nutilde, \Mtilde)}(0), 
\\ \widehat{P}^{(1, 1)}(1),  \ldots, \widehat{P}^{(\Nutilde, \Mtilde)}(1), \ldots, \\
\widehat{P}^{(1, 1)}(Q-1),  \ldots, \widehat{P}^{(\Nutilde, \Mtilde)}(Q-1)]^\trasp
\end{multline}
where
\be
\widehat{P}^{(j, i)}(\widetilde{s}) = \frac{1}{S \, F_i}  
\sum_{\ell=0}^{F_i-1} 
\left \|\overline{\bm y}^{(j, k_{i,\ell})}(\widetilde{s}) \right \|_2^2
\label{eq:object-est}
\ee
for $j \in \{1,2,\ldots,\Nutilde\}$ and $i \in \{1,2,\ldots,\Mtilde\}$, 
within beacon slot $\widetilde{s} \in \{0,1,\ldots, Q-1\}$.

\subsection{Error analysis}
\label{sec:analysis}

As it is apparent from \eqref{eq:p}, the impact of the jamming attack on the BA 
procedure between the BS and the UE is determined by the 
transmit beamforming codebook of the jammer, which appears in 
the matrix $\G_\text{J}$, and the second-order statistics
of the channel 
between the jammer and the UE, i.e., the sparse vector $\xibold_\text{J}$.
The solution of \eqref{eq:nnls} approximates $\xibold_\Bs$ with an  error 
\be 
\mathcal{E}(\widehat{\xibold}_\Bs) \eqdef \|\xibold_\Bs-\widehat{\xibold}_\Bs \|_2
\label{eq:error}
\ee
which  depends not only on the jamming contribution 
$\G_\text{J} \, \xibold_\text{J}$, but also on the fact that 
$\xibold_\Bs$ is not exactly sparse, i.e., only a small
number of its entries are nonzero,
but $\xibold_\Bs$ is only close to a sparse vector.
More precisely, a vector $\s_\Tx \in \Rset^{M_\Tx \Nu}$is called {\em $\kappa_\Tx$-sparse} \cite[Def.~2.1]{Rau} if at most  $\kappa_\Tx$ 
of its entries are nonzero, i.e., $|\text{supp}(\s_\Tx)| \le \kappa_\Tx$,
for $\text{TX} \in \{\text{B}, \text{J}\}$.
The {\em best $\kappa_\Tx$-term approximation} of 
$\xibold_\Tx$ is defined  as (see, e.g., \cite[Def.~2.2]{Rau})
\begin{multline}
\sigma_{\kappa_\Tx}(\xibold_\Tx) \eqdef \inf \Big\{ \| \xibold_\Tx- \s_\Tx \|_1,
\text{where $\s_\Tx \in \Rset^{M_\Tx \Nu}$}
 \\ 
\text{is $\kappa_\Tx$-sparse} \Big\} \: .
\label{eq:bestapprox}
\end{multline}
The infimum is achieved in \eqref{eq:bestapprox} by a $\kappa_\Tx$-sparse vector
$\s_\Tx \in \Cset^{M_\Tx \Nu}$
whose nonzero entries equal the $\kappa_\Tx$ largest absolute entries of $\xibold_\Tx$.
As regards to the transmit beamforming codebook of the jammer, we study 
the two different cases $\G_\text{J} \neq \G_\Bs$ and $\G_\text{J} = \G_\Bs$ separately.

\subsubsection{$\G_\text{J} \neq \G_\Bs$}

In principle, the transmit beamforming codebooks of the BS and
jammer may be different. 
For instance, the jamming 
codebook $\ubtilde^{(i)}_{\text{J}}(\widetilde{s}) $
might be chosen in a pseudo-random manner similarly to the BS
or, if the jammer is a high-power device that has a large amount of power to be spent, 
another option for the jammer could consist of 
probing the channel along 
{\em all} the possible directions  (referred to as {\em omnidirectional beamforming}) and, consequently, setting  
$\ubtilde^{(i)}_{\text{J}}(\widetilde{s}) = \bm{1}_{\Mtilde}/ \sqrt{\Mtilde}$.
In the case of $\G_\text{J} \neq \G_\Bs$, the jamming contribution 
$\G_\text{J} \, \xibold_\text{J}$ appears as additional noise of arbitrary nature
and the reconstruction error \eqref{eq:error} can be upper bounded \cite{Kue} as follows
\be
\mathcal{E}(\widehat{\xibold}_\Bs) \le \frac{A_1}{\sqrt{\kappa_\Bs}} \, \sigma_{\kappa_\Bs}(\xibold_\Bs) + A_2 \, \|\G_\text{J} \, \xibold_\text{J}\|_2
\ee
for some constants $A_1, A_2 >0$, provided that the matrix $\G_\Bs$ satisfies the 
conditions summarized in the Appendix.
By resorting to the sub-multiplicative property of the $\ell_2$ norm \cite{Horn}, one has
\barr
\|\G_\text{J} \, \xibold_\text{J}\|_2 & \le \sqrt{\trace(\G_\text{J} \, \G_\text{J}^\trasp)} 
\, \|\xibold_\text{J}\|_2 
\nonumber \\ & = \frac{\pot_\text{J}}{U_\text{J} \, V} 
\sqrt{
\sum_{\shortstack{\footnotesize
$i \in \{1,2,\ldots,\Mtilde\}$ \\ \footnotesize 
$j \in \{1,2,\ldots,\Nutilde\}$ \\ \footnotesize
$\widetilde{s} \in \{0,1,\ldots, Q-1\}$}
} 
\left \| \bm{1}_{\mathcal{U}_{\text{J}}^{(i)}(\widetilde{s})} \otimes \bm{1}_{\mathcal{V}^{(j)}(\widetilde{s})}\right\|_2^2} 
\nonumber \\ &  \hspace{25mm} \cdot
\sqrt{\sum_{n=1}^{M_\text{J} \Nu} \{\bm R_{\widetilde{\bm C}_\text{J}}\}_{n,n}^2}
\nonumber \\ & \le \frac{\pot_\text{J}}{U_\text{J} \, V} \,
\sqrt{\Mtilde \, \Nutilde \, Q \, M_\text{J} \, \Nu} \, \sqrt{\sum_{n=1}^{M_\text{J} \Nu} \{\bm R_{\widetilde{\bm C}_\text{J}}\}_{n,n}^2}
\label{eq:tracein}
\earr
where we have also used \eqref{eq:gtildesimpl}
and  \eqref{eq:Gtx}, and remembered that 
$\gb^{(j,i)}_{\text{J}}(\widetilde{s}) = \sqrt{U_\text{J}} \, \sqrt{V} \, 
\gbtilde^{(j,i)}_{\text{J}}(\widetilde{s})$.
It is apparent from \eqref{eq:tracein} that probing more directions simultaneously (i.e., increasing $U_\text{J}$) has the 
detrimental effect from the jammer's viewpoint of
spreading the total power over all such directions, thereby  
obtaining a worse power concentration in the angle domain.

\subsubsection{$\G_\text{J} = \G_\Bs$}

The jammer might transmit by using the same beamforming codebook of the BS
that we remember to be known  to all UEs a priori, i.e., 
$\mathcal{U}_{\text{J}}^{(i)}(\widetilde{s}) \equiv \mathcal{U}_{\Bs}^{(i)}(\widetilde{s})$,
for any $i \in \{1,2,\ldots,\Mtilde\}$ 
and $\widetilde{s} \in \{0,1,\ldots, Q-1\}$, which necessarily requires that 
$\Mj=M_\Bs \equiv M$. 
In this case, one has   
$\G_\text{J}=\G_\Bs \equiv \G$ and, consequently, eq.~\eqref{eq:p} ends up to
\be
\bm p  
= \G \, (\xibold_\Bs + \xibold_\text{J}) + \sigma^2_w \, \bm{1}_{\Mtilde \Nutilde Q} 
\label{eq:p-equalcode}
\ee
which shows that the UE sees the sum of two sparse vectors $\xibold_\Bs$ and
$\xibold_\text{J}$ under the same measurement matrix $\G$. This case is 
worse than the previous one when $\G_\text{J} \neq \G_\Bs$ since  
$\widehat{\xibold}_\Bs$ turns out to be an estimate of $\xibold_\Bs+\xibold_\text{J}$. 
In this worst case, successful BA 
between the BS and the UE is 
achieved if 
\be
\frac{\pot_{\Bs}}{\pot_\text{J}} \gg \frac{\max_{n \in \{1,2,\ldots, M\Nu\}} \{\bm R_{\widetilde{\bm C}_\text{J}}\}_{n,n}}{\max_{n \in \{1,2,\ldots, M\Nu\}} \{\bm R_{\widetilde{\bm C}_\Bs}\}_{n,n}} \:.
\label{eq:condequalG}
\ee
Condition \eqref{eq:condequalG} is violated when the jammer transmits with
a power $\pot_\text{J}$ sufficiently greater than $\pot_{\Bs}$ and/or, 
compared to the BS, it has a more favorable propagation towards the UE.

\section{The proposed anti-jamming beam alignment scheme}
\label{sec:random-probing}

In this section, we modify the transmit scheme of the BS in order to allow 
the UE to cancel the jamming contribution.
A key ingredient of our proposed anti-jamming scheme is the
random probing symbols transmitted by the BS, which 
follow the model
\be
d^{(k_{i,\ell})}_{\Bs}(s) = \sqrt{[1-\gamma_{\Bs}(s)] \,\pot_{\Bs}} \, 
t^{(k_{i,\ell})}(s) +
\sqrt{\gamma_{\Bs}(s) \, \pot_{\Bs}} \, 
r^{(k_{i,\ell})}_{\Bs}(s)
\label{eq:d}
\ee
where each stream $\{r^{(k_{i,\ell})}_{\Bs}(s)\}$
is modeled  as a sequence of 
zero-mean unit-variance i.i.d. complex
circular RVs, with 
$r^{(k_{i,\ell})}_{\text{B}}(s)$ and \eqref{eq:d-j} mutually independent 
and statistically independent of noise $\overline{w}^{(j, k_{i,\ell})}(s)$,
for each OFDM block, and for any $i \in \{1,2,\ldots,\Mtilde\}$
and $\forall j \in \{1,2,\ldots,\Nutilde\}$.
The BS allocates a different fraction $0 \le \gamma_\text{B}(s) \le 1$ of $\pot_\text{B}$  
to the random symbols $r^{(k_{i,\ell})}_{\text{B}}(s)$. 
Since $\{r^{(k_{i,\ell})}_{\text{B}}(s)\}$ is randomly generated at 
the BS, it is  unknown at the UE.
However, the UE knows that the BS has superimposed 
the random  sequence $\{r^{(k_{i,\ell})}_{\text{B}}(s)\}$
on the known sequence 
$\{t^{(k_{i,\ell})}(s)\}$ and it can use such a knowledge
to undo the jamming attack. 
The conventional probing scheme \eqref{eq:d-conv}
can be obtained from \eqref{eq:d} 
by setting $\gamma_{\Bs}(s)=0$, $\forall s \in \{0,1,\ldots, W-1\}$.

\begin{figure}[t]
\centering
\includegraphics[width=\columnwidth]{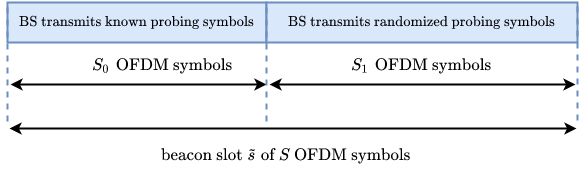}
\caption{Each beacon slot is divided into two subslots: during the first 
$S_0$ OFDM symbols, the BS transmits a known probing sequence, whereas
a random sequence is superimposed to the known symbols in the
remaining $S_1$ OFDM blocks, with $S_0+S_1=S$.
}
\label{fig:fig_3}
\end{figure}

In the sequel, we assume that $\gamma_\text{B}(s)$ does not vary from a 
beacon slot to another, but it might assume different values within a 
beacon slot, i.e, $\gamma_\text{B}(\widetilde{s} \, S + s') \equiv \gamma_\text{B}^{(s')}$,
for any $\widetilde{s} \in \{0,1,\ldots, Q-1\}$ and $s' \in \{0, 1, \ldots, S-1\}$.
To counteract the detrimental effect of the jamming attack, we additionally propose 
to divide each beacon slot $\widetilde{s}$ in two subslots (see Fig.~\ref{fig:fig_3}): in the former one, which lasts 
$S_0$ OFDM symbols, the BS transmits only the known symbols 
$t^{(k_{i,\ell},s')}$ defined in Section \ref{sec:conv}, i.e.,  
$\gamma_\text{B}^{(s')}=0$, for $s' \in \{0, 1, \ldots, S_0-1\}$;
whereas in the remaining $S_1 \eqdef S-S_0$ OFDM symbols of 
each beacon slot, the BS superimposes the random sequence 
$r^{(k_{i,\ell},s')}_{\Bs}(\widetilde{s}) \eqdef 
r^{(k_{i,\ell})}_{\Bs}(\widetilde{s} \, S + s')$
to the known symbols $t^{(k_{i,\ell},s')}$, 
with a fixed power fraction 
$\gamma_\text{B}^{(s')} \equiv \gamma_\text{B} \in (0, 1]$,
for all $s' \in \{S_0, S_0+1, \ldots, S-1\}$.

By substituting \eqref{eq:d-j} and \eqref{eq:d} into \eqref{eq:overline_y}, one has
\begin{multline}
\overline{y}^{(j, k_{i,\ell},s')}(\widetilde{s})  =
\htilde^{(j, i,  k_{i,\ell})}_{\Bs}(\widetilde{s}) 
\left\{\sqrt{[1-\gamma_\text{B}^{(s')}] \,\pot_\text{B}} \, t^{(k_{i,\ell}, s')}(\widetilde{s})
\right. \\ \left.
 + \sqrt{\gamma_\text{B}^{(s')} \,\pot_\text{B}} \, r_{\Bs}^{(k_{i,\ell}, s')}(\widetilde{s})\right\}
\\
+ \htilde^{(j, i,  k_{i,\ell})}_{\text{J}}(\widetilde{s}) 
\left\{\sqrt{(1-\gamma_\text{J}) \,\pot_\text{J}} \, t^{(k_{i,\ell}, s')}(\widetilde{s})
\right. \\ \left.
+ \sqrt{\gamma_\text{J}\,\pot_\text{J}} \, r^{(k_{i,\ell}, s')}_{\text{J}}(\widetilde{s})\right\}
+ \overline{w}^{(j, k_{i,\ell},s')}(\widetilde{s})
\label{eq:overline_y_mod}
\end{multline}
where 
$\widetilde{s} \in \{0,1,\ldots, Q-1\}$ and $s' \in \{0, 1, \ldots, S-1\}$, 
with
$\htilde^{(j, i,  k_{i,\ell})}_{\Tx}(\widetilde{s}) $, 
and 
$r^{(k_{i,\ell},s')}_{\text{J}}(\widetilde{s})$
defined in Section~\ref{sec:conv}.  
According to the proposed protocol, the data block \eqref{eq:yvec}
received by the UE  during the beacon slot $\widetilde{s} \in \{0,1,\ldots, Q-1\}$
can be partitioned as 
\begin{multline}
\overline{\bm y}^{(j, k_{i,\ell})}(\widetilde{s})=
\begin{bmatrix}
\overline{\bm y}_0^{(j, k_{i,\ell})}(\widetilde{s})
\\
\overline{\bm y}_1^{(j, k_{i,\ell})}(\widetilde{s})]
\end{bmatrix} \:,  \\
\text{with $\overline{\bm y}_0^{(j, k_{i,\ell})}(\widetilde{s}) \in \Cset^{S_0}$
and $\overline{\bm y}_1^{(j, k_{i,\ell})}(\widetilde{s}) \in \Cset^{S_1}$}
\label{eq:part}
\end{multline}
for $j \in \{1,2,\ldots,\Nutilde\}$, $i \in \{1,2,\ldots,\Mtilde\}$, 
and beacon slot $\widetilde{s} \in \{0,1,\ldots, Q-1\}$, with 
\begin{multline}
\overline{\bm y}_0^{(j, k_{i,\ell})}(\widetilde{s}) = \sqrt{\pot_\text{B} } \,  
\htilde^{(j, i,  k_{i,\ell})}_{\Bs}(\widetilde{s}) \, \bm t_0^{(k_{i,\ell})}(\widetilde{s}) 
\\ + \htilde^{(j, i,  k_{i,\ell})}_{\text{J}}(\widetilde{s}) 
\left\{\sqrt{(1-\gamma_\text{J}) \,\pot_\text{J}} \, \bm t_0^{(k_{i,\ell})}(\widetilde{s})
\right. \\ \left.
+ \sqrt{\gamma_\text{J}\,\pot_\text{J}} \, \bm r_{0,\text{J}}^{(k_{i,\ell})}(\widetilde{s})
\right\} + \overline{\bm w}_0^{(j, k_{i,\ell})}(\widetilde{s})
\label{eq:yvec-0}
\end{multline}
\begin{multline}
\overline{\bm y}_1^{(j, k_{i,\ell})}(\widetilde{s}) =   
\htilde^{(j, i,  k_{i,\ell})}_{\Bs}(\widetilde{s}) 
\left\{\sqrt{(1-\gamma_\Bs) \,\pot_\Bs} \, \bm t_1^{(k_{i,\ell})}(\widetilde{s})
\right. \\ \left.
+ \sqrt{\gamma_\Bs\,\pot_\Bs} \, \bm r_{1,\text{B}}^{(k_{i,\ell})}(\widetilde{s})
\right\}
+ \htilde^{(j, i,  k_{i,\ell})}_{\text{J}}(\widetilde{s}) 
\left\{\sqrt{(1-\gamma_\text{J}) \,\pot_\text{J}} \, \bm t_1^{(k_{i,\ell})}(\widetilde{s})
\right. \\ \left. + \sqrt{\gamma_\text{J}\,\pot_\text{J}} \, \bm r_{1,\text{J}}^{(k_{i,\ell})}(\widetilde{s})
\right\} + \overline{\bm w}_1^{(j, k_{i,\ell})}(\widetilde{s})
\label{eq:yvec-1}
\end{multline}
where $\bm t_0^{(k_{i,\ell})}(\widetilde{s}) \in \Cset^{S_0}$, 
$\bm t_1^{(k_{i,\ell})}(\widetilde{s}) \in \Cset^{S_1}$, 
$\bm r_{0,\text{J}}^{(k_{i,\ell})}(\widetilde{s}) \in \Cset^{S_0}$, 
$\bm r_{1,\text{J}}^{(k_{i,\ell})}(\widetilde{s}) \in \Cset^{S_1}$, 
$\overline{\bm w}_0^{(j, k_{i,\ell})}(\widetilde{s}) \in \Cset^{S_0}$,
$\overline{\bm w}_1^{(j, k_{i,\ell})}(\widetilde{s}) \in \Cset^{S_1}$
are obtained by partitioning 
$\bm t^{(k_{i,\ell})}(\widetilde{s})$,
$\bm r_{\text{J}}^{(k_{i,\ell})}(\widetilde{s})$,
and $\overline{\bm w}^{(j, k_{i,\ell})}(\widetilde{s})$,
respectively, in accordance with \eqref{eq:part}, and, moreover, 
\begin{multline}
\bm r_{1,\text{B}}^{(k_{i,\ell})}(\widetilde{s}) \eqdef [r_{\text{B}}^{(k_{i,\ell},S_0)}(\widetilde{s}), r_{\text{B}}^{(k_{i,\ell},S_0+1)}(\widetilde{s}), 
\\ \ldots, r_{\text{B}}^{(k_{i,\ell},S-1)}(\widetilde{s})]^{\trasp}
\in \Cset^{S_1} \: .
\end{multline}

Starting from \eqref{eq:yvec-0}-\eqref{eq:yvec-1}, 
accordingly to the modified transmit 
protocol of the BS depicted in Fig.~\ref{fig:fig_3}, 
we additionally propose to modify
the BA procedure implemented by the UE.
In our proposal, the UE performs BA in three steps.
In the first step,  for any beacon slot, 
the received blocks $\overline{\bm y}_0^{(j, k_{i,\ell})}(\widetilde{s})$
and $\overline{\bm y}_1^{(j, k_{i,\ell})}(\widetilde{s})$ are projected
onto the subspace that is orthogonal 
to the subspace generated by the 
corresponding known probing symbols.
In the second step, the power of 
the jammer-plus-noise contribution is estimated in each beacon slot
by processing the projected version of $\overline{\bm y}_0^{(j, k_{i,\ell})}(\widetilde{s})$.
In the last step, the BA procedure is finalized by 
using the 
projected version of $\overline{\bm y}_1^{(j, k_{i,\ell})}(\widetilde{s})$,
for each beacon slot, thus developing a ``cleaned'' NNLS optimization problem
that is obtained by 
canceling out the previously estimated jammer-plus-noise 
power contribution.

\subsection{Step 1: Subspace projections}
\label{sec:step-1}

Both the power estimation of the jammer-plus-noise contribution obtained from 
$\overline{\bm y}_0^{(j, k_{i,\ell})}(\widetilde{s})$
and the BA algorithm applied on $\overline{\bm y}_1^{(j, k_{i,\ell})}(\widetilde{s})$ are performed 
in the subspace that is orthogonal 
to the one-dimensional subspace generated by the 
known vectors $\bm t_0^{(k_{i,\ell})}(\widetilde{s})$ and 
$\bm t_1^{(k_{i,\ell})}(\widetilde{s})$, respectively.
Specifically, for $\kappa \in \{0,1\}$, let 
$\mathcal{\bm{P}}_{\bm t_\kappa^{(k_{i,\ell})}(\widetilde{s})}^{\perp} \in \Cset^{S_\kappa 
\times S_\kappa}$ 
denote the orthogonal projector onto the subspace complementary to that 
spanned by $\bm t_\kappa^{(k_{i,\ell})}(\widetilde{s})$, 
it results that
\be
\mathcal{\bm{P}}_{\bm t_\kappa^{(k_{i,\ell})}(\widetilde{s})}^{\perp} 
= \I_{S_\kappa}-\frac{1}{S_\kappa} \, \bm t_\kappa^{(k_{i,\ell})}(\widetilde{s}) 
\, [\bm t_\kappa^{(k_{i,\ell})}(\widetilde{s})]^{\herm}
\ee
where we have used the fact that $\|\bm t_\kappa^{(k_{i,\ell})}(\widetilde{s})\|_2^2 =S_\kappa$. 
By construction the matrix $\mathcal{\bm{P}}_{\bm t_\kappa^{(k_{i,\ell})}(\widetilde{s})}^{\perp} 
$ has rank equal to $S_\kappa-1$.
Therefore, the economy-size eigenvalue decomposition (EVD) of 
$\mathcal{\bm{P}}_{\bm t_\kappa^{(k_{i,\ell})}(\widetilde{s})}^{\perp}$ is given by
$\mathcal{\bm{P}}_{\bm t_\kappa^{(k_{i,\ell})}(\widetilde{s})}^{\perp} 
= \bm U_\kappa^{(k_{i,\ell})}(\widetilde{s}) \, \bm \Sigma_\kappa^{(k_{i,\ell})}(\widetilde{s}) \, 
[\bm U_\kappa^{(k_{i,\ell})}(\widetilde{s})]^\herm$, 
where $\bm U_\kappa^{(k_{i,\ell})}(\widetilde{s}) \in \Cset^{S_\kappa \times (S_\kappa-1)}$ represents the semi-unitary 
eigenvector matrix, obeying $[\bm U_\kappa^{(k_{i,\ell})}(\widetilde{s})]^\herm \, 
\bm U_\kappa^{(k_{i,\ell})}(\widetilde{s}) = \I_{S_\kappa-1}$, 
whereas the diagonal matrix $\bm \Sigma_\kappa^{(k_{i,\ell})}(\widetilde{s}) \in \Rset^{(S_\kappa-1) \times (S_\kappa-1)}$
contains the nonzero eigenvalues of $\mathcal{\bm{P}}_{\bm t_\kappa^{(k_{i,\ell})}(\widetilde{s})}^{\perp}$. 

The part of the BS and jamming contribution associated with the transmission of the known probing symbols can be canceled out by applying the 
linear operator $[\bm U_\kappa^{(k_{i,\ell})}(\widetilde{s})]^\herm$ 
on $\overline{\bm y}_\kappa^{(j, k_{i,\ell})}(\widetilde{s})$ given by 
\eqref{eq:yvec-0}-\eqref{eq:yvec-1}, for $\kappa \in \{0,1\}$, thus yielding 
\barr
\overline{\bm y}_{0,\perp}^{(j, k_{i,\ell})}(\widetilde{s}) & \eqdef
[\bm U_0^{(k_{i,\ell})}(\widetilde{s})]^\herm \, \overline{\bm y}_0^{(j, k_{i,\ell})}(\widetilde{s}) 
\nonumber \\ & = 
\sqrt{\gamma_\text{J} \, \pot_\text{J}} \, \htilde^{(j, i,  k_{i,\ell})}_{\text{J}}(\widetilde{s}) \, 
[\bm U^{(k_{i,\ell})}_0(\widetilde{s})]^\herm \, \bm r_{0,\text{J}}^{(k_{i,\ell})}(\widetilde{s})
\nonumber \\ &
+ [\bm U^{(k_{i,\ell})}_0(\widetilde{s})]^\herm \, \overline{\bm w}_0^{(j, k_{i,\ell})}(\widetilde{s}) 
\label{eq:yvec-00}
\\
\overline{\bm y}_{1,\perp}^{(j, k_{i,\ell})}(\widetilde{s}) & \eqdef
[\bm U_1^{(k_{i,\ell})}(\widetilde{s})]^\herm \, \overline{\bm y}_1^{(j, k_{i,\ell})}(\widetilde{s}) 
\nonumber \\ & = 
\sqrt{\gamma_\text{B} \, \pot_\text{B}} \, \htilde^{(j, i,  k_{i,\ell})}_{\text{B}}(\widetilde{s}) \, 
[\bm U^{(k_{i,\ell})}_1(\widetilde{s})]^\herm \, \bm r_{1,\text{B}}^{(k_{i,\ell})}(\widetilde{s})
\nonumber \\ & +
\sqrt{\gamma_\text{J} \, \pot_\text{J}} \, \htilde^{(j, i,  k_{i,\ell})}_{\text{J}}(\widetilde{s}) \, 
[\bm U^{(k_{i,\ell})}_1(\widetilde{s})]^\herm \, \bm r_{1,\text{J}}^{(k_{i,\ell})}(\widetilde{s})
\nonumber \\ &
+ [\bm U^{(k_{i,\ell})}_1(\widetilde{s})]^\herm \, \overline{\bm w}_1^{(j, k_{i,\ell})}(\widetilde{s})
\label{eq:yvec-11}
\earr
for $j \in \{1,2,\ldots,\Nutilde\}$, $i \in \{1,2,\ldots,\Mtilde\}$, 
and beacon slot $\widetilde{s} \in \{0,1,\ldots, Q-1\}$.

The projected vector $\overline{\bm y}_{0,\perp}^{(j, k_{i,\ell})}(\widetilde{s})$ - from which 
the BS contribution has been removed - is used in Step~2 to estimate the jammer-plus-noise
power, whereas the projected vector $\overline{\bm y}_{1,\perp}^{(j, k_{i,\ell})}(\widetilde{s})$
is the input of the BA procedure in Step~3.

\subsection{Step 2: Estimation of the jammer-plus-noise contribution}
\label{sec:step-2}

Having removed the BS contribution from the received data in the 
first part of each beacon slot, it is now possible to estimate from \eqref{eq:yvec-00}
the power of 
the jammer-plus-noise term 
at the output of the $j$-th RF chain of the UE
due to the signal transmitted by
$i$-th RF chain of the jammer 
in the $\widetilde{s}$-th beacon slot
through the estimator
\be
P^{(j, i)}_{0,\perp}(\widetilde{s}) =\frac{1}{(S_0-1) \, F_i}  
\sum_{\ell=0}^{F_i-1} 
\Es\left[\left \|\overline{\bm y}_{0,\perp}^{(j, k_{i,\ell})}(\widetilde{s}) \right \|_2^2\right] 
\label{eq:object-00}
\ee
for $j \in \{1,2,\ldots,\Nutilde\}$, $i \in \{1,2,\ldots,\Mtilde\}$, 
and beacon slot $\widetilde{s} \in \{0,1,\ldots, Q-1\}$, 
where the expectation is also evaluated with respect to the random probing
symbols transmitted by the jammer.
Under our assumptions \eqref{eq:object-00} can be explicated as 
\be
P^{(j, i)}_{0,\perp}(\widetilde{s}) = \gamma_\text{J} \,  [\gb^{(j,i)}_\text{J}(\widetilde{s})]^\trasp \, 
\xibold_\text{J} + \sigma^2_w
\label{eq:object-00-two}
\ee
where we have used \eqref{eq:Eh}
and the facts that 
\barr
\Es \left[ \|
[\bm U^{(k_{i,\ell})}_\kappa(\widetilde{s})]^\herm \, \bm r_{\kappa,\text{J}}^{(k_{i,\ell})}(\widetilde{s})
\|_2^2 \right] & = (S_\kappa-1) 
\nonumber 
\\
\Es \left[ \|[\bm U^{(k_{i,\ell})}_\kappa(\widetilde{s})]^\herm \, \overline{\bm w}_\kappa^{(j, k_{i,\ell})}(\widetilde{s})\|_2^2 \right] & = (S_\kappa-1) \, 
\sigma^2_w
\nonumber
\earr
for $\kappa \in \{0,1\}$, 
due to the semi-unitary property of $\bm U^{(k_{i,\ell})}_\kappa(\widetilde{s})$.
The $P^{(j, i)}_{0,\perp}(\widetilde{s})$ also includes the noise variance 
$\sigma_w^2$, whose knowledge is thereby not required for beam determination. 
In practice, the power level $P^{(j, i)}_{0,\perp}(\widetilde{s})$
can be directly estimated from data as 
\be
\widehat{P}^{(j, i)}_{0,\perp}(\widetilde{s})= \frac{1}{(S_0-1) \, F_i}  
\sum_{\ell=0}^{F_i-1} 
\left \|\overline{\bm y}_{0,\perp}^{(j, k_{i,\ell})}(\widetilde{s}) \right \|_2^2 \:. 
\label{eq:object-00-est}
\ee
The obtained power estimates \eqref{eq:object-00-est} are used in Step~3 to achieve the BA 
between the BS and the UE in an optimization process that is (nearly) free from 
the jammer-plus-noise contribution.

\subsection{Step 3: BA with jammer-plus-noise cancellation}
\label{sec:step-3}

The BA process is based on \eqref{eq:yvec-11} and exploits the power estimations 
provided in the previous step. Similarly to \eqref{eq:object}, the NNLS optimization process relies on the power measurements
\barr
P^{(j, i)}_{1,\perp}(\widetilde{s}) & = \frac{1}{(S_1-1) \, F_i}  
\sum_{\ell=0}^{F_i-1} 
\Es\left[\left \|\overline{\bm y}_{1,\perp}^{(j, k_{i,\ell})}(\widetilde{s}) \right \|_2^2\right] 
\nonumber \\ & =
\gamma_\text{B} \, [\gb^{(j,i)}_\text{B}(\widetilde{s})]^\trasp \, \xibold_\text{B} + \pot^{(j, i)}_{0,\perp}(\widetilde{s})
\label{eq:object-1}
\earr
for $j \in \{1,2,\ldots,\Nutilde\}$, $i \in \{1,2,\ldots,\Mtilde\}$, 
and beacon slot $\widetilde{s} \in \{0,1,\ldots, Q-1\}$, where 
the expectation is also evaluated with respect to the random probing
symbols transmitted by both the BS and the jammer, and
the equality 
follows from arguments similar to those invoked in 
Sections~\ref{sec:conv} and \ref{sec:step-2}, with the additional observation
that $\Es \left[ \|
[\bm U^{(k_{i,\ell})}_1(\widetilde{s})]^\herm \, \bm r_{1,\text{B}}^{(k_{i,\ell})}(\widetilde{s})
\|_2^2 \right] = (S_1-1) $ and
$P^{(j, i)}_{0,\perp}(\widetilde{s})$ is given by 
\eqref{eq:object-00-two}.
By defining 
\begin{multline}
\bm p_{\kappa}^{\perp}  \eqdef [P_{\kappa,\perp}^{(1, 1)}(0),  \ldots, 
P_{\kappa,\perp}^{(\Nutilde, \Mtilde)}(0), 
\\ P_{\kappa,\perp}^{(1, 1)}(1),  \ldots, 
P_{\kappa,\perp}^{(\Nutilde, \Mtilde)}(1), \ldots, \\ 
P_{\kappa,\perp}^{(1, 1)}(Q-1),  \ldots, P_{\kappa,\perp}^{(\Nutilde, \Mtilde)}(Q-1)]^\trasp
\label{eq:pperp}
\end{multline}
for $\kappa \in \{0,1\}$, one gets the vector model
\be
\bm p_{1}^{\perp} = \G_\Bs \, \xibold_\Bs^{\perp} + 
\underbrace{\G_\text{J} \, \xibold_\text{J}^{\perp} + \sigma^2_w \, \bm{1}_{\Mtilde \Nutilde Q}}_{\bm p_{0}^{\perp}} = \G_\Bs \, \xibold_\Bs^{\perp} + \bm p_{0}^{\perp}
\label{eq:pperp-1}
\ee
with $\xibold_\Tx^{\perp} \eqdef \gamma_\Tx \, \xibold_\Tx \in \Rset^{\Mtx \Nu}$, where 
$\xibold_\Tx$ and $\G_\Tx$ are defined by 
\eqref{xibold} and \eqref{eq:Gtx}, respectively.
In the proposed anti-jamming BA procedure, 
the sparse vector $\xibold_\Bs^{\perp}$ can be reconstructed 
from the measurements of the form \eqref{eq:pperp-1}
via the {\em modified} NNLS optimization problem:
\begin{multline}
\widehat{\xibold}_\Bs^{\perp}=\arg \min_{\xibold_\Bs^\star \in \Rset^{M_\Bs \Nu}}
\left \|\bm p_{1}^{\perp} - \G_\Bs \, \xibold_\Bs^\star - \bm p_{0}^{\perp}\right\|_2^2\: ,
\\ \text{subject to $\xibold_\Bs^\star \ge \zerobd_{M_\Bs \Nu}$}
\label{eq:nnls-perp}
\end{multline}
for which the algorithm of Lawson
and Hanson is particularly well adapted \cite{Bjorck}.
Strictly speaking, the effect of the jamming attack is counteracted by 
subtracting the contribution of the jammer-plus-noise 
from the received power. 
Practical implementation of the proposed NNLS problem mandates the replacement of 
$\bm p_{\kappa}^{\perp}$
in \eqref{eq:nnls-perp} with 
\begin{multline}
\widehat{\bm p}_{\kappa}^{\perp}  \eqdef [\widehat{P}_{\kappa,\perp}^{(1, 1)}(0),  \ldots, \widehat{P}_{\kappa,\perp}^{(\Nutilde, \Mtilde)}(0), 
\\ \widehat{P}_{\kappa,\perp}^{(1, 1)}(1),  \ldots, \widehat{P}_{\kappa,\perp}^{(\Nutilde, \Mtilde)}(1), \ldots, \\
\widehat{P}_{\kappa,\perp}^{(1, 1)}(Q-1),  \ldots, \widehat{P}_{\kappa,\perp}^{(\Nutilde, \Mtilde)}(Q-1)]^\trasp
\end{multline}
for $\kappa \in \{0,1\}$, 
where $\widehat{P}^{(j, i)}_{0,\perp}(\widetilde{s})$ has been defined in
\eqref{eq:object-00-est} and
\be
\widehat{P}^{(j, i)}_{1,\perp}(\widetilde{s}) = \frac{1}{(S_1-1) \, F_i}  
\sum_{\ell=0}^{F_i-1} 
\left \|\overline{\bm y}_{1,\perp}^{(j, k_{i,\ell})}(\widetilde{s}) \right \|_2^2  
\label{eq:object-11-est}
\ee
for $j \in \{1,2,\ldots,\Nutilde\}$, $i \in \{1,2,\ldots,\Mtilde\}$, 
and beacon slot $\widetilde{s} \in \{0,1,\ldots, Q-1\}$.

\subsection{Remarks}
\label{sec:remark}

Some remarks are now in order regarding the proposed anti-jamming 
BA approach.

\vspace{3mm}
{\em Remark~1:}
Our general framework allows us to consider different jamming attacks. 
If the jammer transmits only known probing symbols, i.e., 
$\gamma_\text{J}=0$ in \eqref{eq:d-j}, its contribution disappears
from the projected data $\overline{\bm y}_{0,\perp}^{(j, k_{i,\ell})}(\widetilde{s})$
and $\overline{\bm y}_{1,\perp}^{(j, k_{i,\ell})}(\widetilde{s})$, 
since the projections are performed onto the subspaces that are orthogonal 
to those spanned by the known probing vectors $\bm t_0^{(k_{i,\ell})}(\widetilde{s})$
and $\bm t_1^{(k_{i,\ell})}(\widetilde{s})$,
for $\widetilde{s} \in \{0,1,\ldots, Q-1\}$.
In this type of attack, 
the procedure in Step~2 provides estimation of the noise variance $\sigma_w^2$ only
and the BA algorithm in Step~3 operates in a jammer-free scenario. 
On the other hand, when the jammer adds noise to the known probing symbols, i.e., 
$0 < \gamma_\text{J} \le 1$ in \eqref{eq:d-j}, the jammer also transmits into
the subspace complementary to those generated by $\bm t_0^{(k_{i,\ell})}(\widetilde{s})$ and
$\bm t_1^{(k_{i,\ell})}(\widetilde{s})$.
In such an adversarial attack,  the jammer-plus-noise power 
is estimated in Step~2 and, then, it is subtracted in Step~3. 
The impact of $\gamma_\text{J}$ on the performance of the proposed anti-jamming 
BA scheme is studied in Section~\ref{sec:simul} (see Tab.~\ref{tab:gamma_j}).

\vspace{3mm}
{\em Remark~2:}
A distinguished feature of our BA technique is that neither 
a preventive detection of the jamming attack nor knowledge 
of the type of attack is required. Indeed, the  proposed BA 
procedure successfully works even in the absence of
the jammer. Such a case is akin to the previously discussed one when 
the jammer transmits known probing symbols only.  

\vspace{3mm}
{\em Remark~3:}
In the proposed BA procedure, the power
transmitted by the BS in the subspace spanned by the known
probing symbols is not used in Step~2 (see also Fig.~\ref{fig:fig_3}),
thus implying a possible waste of energy. 
One can argue that, in principle, the BS could not transmit in the first 
$S_0$ OFDM symbols of each beacon slot by powering-down
its power amplifier(s). So doing, estimation of the jammer-plus-noise power 
in Step~2 could be obtained without performing the subspace projection at the UE.
However, this option may not be feasible in practice for two basic reasons.
First, current 3GPP specifications mandates the use of a continuous transmission
during the beam-sweeping phase \cite{3GPP-2018-6}.
Second, the BS can enter a sleep mode with zero time 
delay; vice versa, going  back  from  a  sleep  mode  to  the  active  
transmission mode requires a certain delay and a certain 
amount of energy, which both depend on the sleep level. 
If the sleep level is arbitrarily  close  to  
zero, a somewhat reduced
power saving may achieved and, moreover, 
the activation process of the BS  might 
require an acceptable wake up time \cite{Fre}.  

\vspace{3mm}
{\em Remark~4:}
Similarly to Step~2, the known part of the probing
signal transmitted by the BS is not exploited for beam determination
in Step~3. 
Henceforth, one might set $\gamma_\Bs=1$ in \eqref{eq:yvec-1}
in order not to squander energy at the BS:
in this case, the BS transmits only
random probing variables during the last 
$S_1$ OFDM symbols of each beacon slot 
(see again Fig.~\ref{fig:fig_3}).
However, the optimal choice of $\gamma_\Bs$ might also be
dictated by other practical constraints, such as 
hardware complexity and impairments \cite{Zaidi}, as well as
compliance with applicable standards,    
codes, and regulations.
It is numerically shown in Section~\ref{sec:simul} 
(see Tab.~\ref{tab:gamma_b}) that values
of $\gamma_\Bs$ slightly smaller than one do not significantly affect
the performance of the proposed anti-jamming BA scheme.

\section{Numerical results}
\label{sec:simul}

In this section, we provide numerical results aimed at evaluating the performance of the 
proposed jamming-resistant beam alignment technique. We consider an OFDM
system, employing $F = 2048$ subcarriers and cyclic prefix of length $L_{\text{cp}} = 128$.
The system operates with carrier frequency $f_0 = 70$ GHz and bandwidth $1/T_\text{c} = 1$ GHz.
We assume that both the BS and jammer 
have $M_{\Bs} = M_{\J} = 32$ antennas and $\widetilde{M} = 3$ RF chains, and the UE has 
$N_{\Ue} = 32$ antennas and $\widetilde{N}_{\Ue} = 2$ RF chains.
The number of subcarriers assigned to each probing stream is constant, i.e., $F_i = 3$, $\forall i \in \{1,2,\ldots, \widetilde{M}\}$. 
The beacon slot contains $S = 28$ OFDM symbols. 
The number of paths of the BS-to-UE and jammer-to-UE links are fixed to $L_{\Bs} = L_{J} = 2$. 
The channel gains $\rho_\Tx(\ell)$, for $\ell \in \{1, 2\}$ and $\Tx \in \{\Bs, \J\}$,
are generated as circularly-symmetric statistically independent complex Gaussian RVs, 
with variance $\sigma^2(\ell)$ independent of TX, for $\ell \in \{1, 2\}$,
and $\sigma^2(1) = 1$ and $\sigma^2(2)$ 3dB less. The delays $\tau_\Tx(\ell)$ are randomly
generated according to the one-sided exponentially decreasing delay power spectrum, i.e., 
$\tau_\Tx(\ell) = -\tau_{\text{slope}} \, \mathrm{ln}[1-u_\ell (1-e^{-\Delta_\ell/\tau_{\text{slope}}})]$,
where the maximum delay $\Delta_\Bs = \Delta_\J=3$ and slop-time $\tau_\text{slope}=2$
(normalized to the sampling period), and $u_k$ are independent RVs uniformly distributed
in the interval $(0,1)$. The AoAs and AoDs of both the BS and jammer are generated as independent RVs uniformly distributed 
into $(-\pi/2,\pi/2)$.
The beamforming codebooks of the BS and UE are chosen 
in a pseudo-random manner as explained in Subsection~\ref{sec:codebook}, with cardinality 
$U_\Bs=4$ and  $V=4$, respectively.
The signal-to-jamming ratio (SJR) is defined as $\text{SJR} \eqdef \pot_\Bs/\pot_\text{J}$.
Unless otherwise specified, the number of beacon slots is $Q=100$
and we set $\gamma_\Bs=1$ (i.e., the BS transmits only
random probing variables during the last $S_1$ symbols of each beacon slot), 
$\gamma_\text{J}=1$ (i.e., the jammer transmits noise only), and $S_0=S_1=14$ (i.e., each beacon
slot is divided in two equal parts), and $\text{SJR}=-5$ dB.

In all the subsequent experiments, we consider three different cases 
regarding the choice of the transmit beamforming codebook of the jammer:
\begin{enumerate}[{Case} 1:]
\itemsep = 0mm
\item
For $\widetilde{s} \in \{0,1,\ldots, Q-1\}$, 
the jamming codebook $\ubtilde^{(i)}_{\text{J}}(\widetilde{s}) $
is chosen in a pseudo-random manner, 
independently of the  
the BS and UE codebooks, with $U_\text{J}=U_\Bs=V=4$. 
\item
The jammer carries out omnidirectional beamforming by 
probing the channel along 
all the possible directions, i.e.,   
$\ubtilde^{(i)}_{\text{J}}(\widetilde{s}) = \bm{1}_{\Mtilde}/ \sqrt{\Mtilde}$.
\item
For any $i \in \{1,2,\ldots,\Mtilde\}$ 
and $\widetilde{s} \in \{0,1,\ldots, Q-1\}$, 
the jammer transmits by using the same beamforming codebook of the BS, i.e.,  $\mathcal{U}_{\text{J}}^{(i)}(\widetilde{s}) \equiv \mathcal{U}_{\Bs}^{(i)}(\widetilde{s})$.
\end{enumerate}

We implement the jammer-unaware BA strategy based on \eqref{eq:nnls} 
and the proposed anti-jamming BA procedure based on \eqref{eq:nnls-perp}.
As an ideal reference, we also report the performance of NNLS BA in the 
absence of the jamming attack by assuming perfect knowledge of the noise 
power $\sigma_w^2$, which is referred to as ``w/o jamming".
As a performance metric, we evaluate the {\em probability $P_\text{BA}$ of successful BA},
which is defined as the probability that the index of the 
largest component of $\widehat{\xibold}_\Bs$ [resp. $\widehat{\xibold}_\Bs^{\perp}$]
coincides with the index of 
the actual largest entry of 
$\xibold_\Bs$ [resp. $\xibold_\Bs^{\perp}$].
In each Monte Carlo run, a new set of random probing symbols, random codebooks, 
noise, and channel parameters 
is randomly generated. The number of 
Monte Carlo runs is $1000$ in all the experiments.

\begin{table}[t]
\centering{}%
\begin{tabular}{lccccccccccc}
\noalign{\vskip\doublerulesep}
& \phantom{abc}  & \multicolumn{7}{c}{${\gamma_{\text{J}}}$} \\
\toprule
$\text{P}_\text{BA}$ & $0.0$ & $0.1$ & $0.2$ & $0.3$ 
& $0.4$ & $0.5$ & $0.6$ 
& $0.7$ & $0.8$ & $0.9$ & $1.0$ \tabularnewline[\doublerulesep]
\toprule
Case 1  &
$0.889$ &  $0.886$ & $0.885$ &$0.885$ &  $0.885$ & $0.884$ & $0.881$ &  $0.880$ & $0.879$ & $0.879$ & $0.878$
\tabularnewline[\doublerulesep]
\noalign{\vskip\doublerulesep}
Case 2 &
$0.890$ &  $0.887$ & $0.884$ & $0.884$ &  $0.884$ & $0.883$ & $0.879$ &  $0.878$ & $0.876$ & $0.876$ & $0.875$
\tabularnewline[\doublerulesep]
\noalign{\vskip\doublerulesep}
Case 3 &
$0.890$ &  $0.885$ & $0.883$ & $0.881$ &  $0.879$ & $0.879$ & $0.873$ &  $0.870$ & $0.866$ & $0.865$ & $0.864$
\tabularnewline[\doublerulesep]
\toprule
\end{tabular}
\caption{$P_\text{BA}$ versus $\gamma_{\text{J}}$ ($\gamma_\Bs=1$, $Q = 100$, and 
$\text{SJR}=-5 \, \mathrm{dB}$).}
\label{tab:gamma_j}
\end{table}
\begin{table}[t]
\centering{}%
\begin{tabular}{lcccccccccc}
\noalign{\vskip\doublerulesep}
& \phantom{abc}  & \multicolumn{7}{c}{${\gamma_{\text{B}}}$} \\
\toprule
$\text{P}_\text{BA}$ & $0.1$ & $0.2$ & $0.3$ 
& $0.4$ & $0.5$ & $0.6$ 
& $0.7$ & $0.8$ & $0.9$ & $1.0$ \tabularnewline[\doublerulesep]
\toprule
Case 1  &
$0.801$ &  $0.840$ & $0.856$ &$0.863$ &  $0.869$ & $0.878$ & $0.882$ &  $0.884$ & $0.884$ & $0.886$
\tabularnewline[\doublerulesep]
\noalign{\vskip\doublerulesep}
Case 2  &
$0.789$ &  $0.829$ & $0.854$ & $0.868$ &  $0.872$ & $0.877$ & $0.879$ &  $0.881$ & $0.882$ & $0.884$
\tabularnewline[\doublerulesep]
\noalign{\vskip\doublerulesep}
Case 3 &
$0.758$ &  $0.807$ & $0.826$ & $0.839$ &  $0.852$ & $0.859$ & $0.865$ &  $0.867$ & $0.870$ & $0.871$
\tabularnewline[\doublerulesep]
\toprule
\end{tabular}
\caption{$P_\text{BA}$ versus $\gamma_{\text{B}}$ ($\gamma_\text{J}=1$, $Q = 100$,
and $\text{SJR}=-5 \, \mathrm{dB}$).}
\label{tab:gamma_b}
\end{table}

\subsection{Probability of successful BA versus $\gamma_\text{J}$ and $\gamma_\Bs$}
\label{sec:1}
Tabs.~\ref{tab:gamma_j} and \ref{tab:gamma_b} report the BA performance of the
proposed procedure as a function of
$\gamma_\text{J}$ and $\gamma_\Bs$, respectively. 
The proposed anti-jamming BA scheme is slightly influenced by the way in which 
the jammer splits its available power between known probing symbols and intentional noise.
We remember that, when $\gamma_\text{J}=0$, i.e., the jammer transmits only known probing symbols, 
the jamming contribution is completely rejected
via orthogonal projection. Therefore, the fact that the performance does not appreciably vary 
for $\gamma_\text{J} > 0$ indirectly corroborates the satisfactory jamming rejection capabilities of
the proposed modified NNLS optimization problem.  
On the other hand, as expected, the optimal value of $\gamma_\Bs$ is equal to one. However,
values of $\gamma_\Bs$ slightly smaller than one lead to a negligible performance degradation. 

\subsection{Probability of successful BA versus $S_0$}
\label{sec:2}
\begin{figure}[t!]
\centering
\includegraphics[width=\columnwidth]{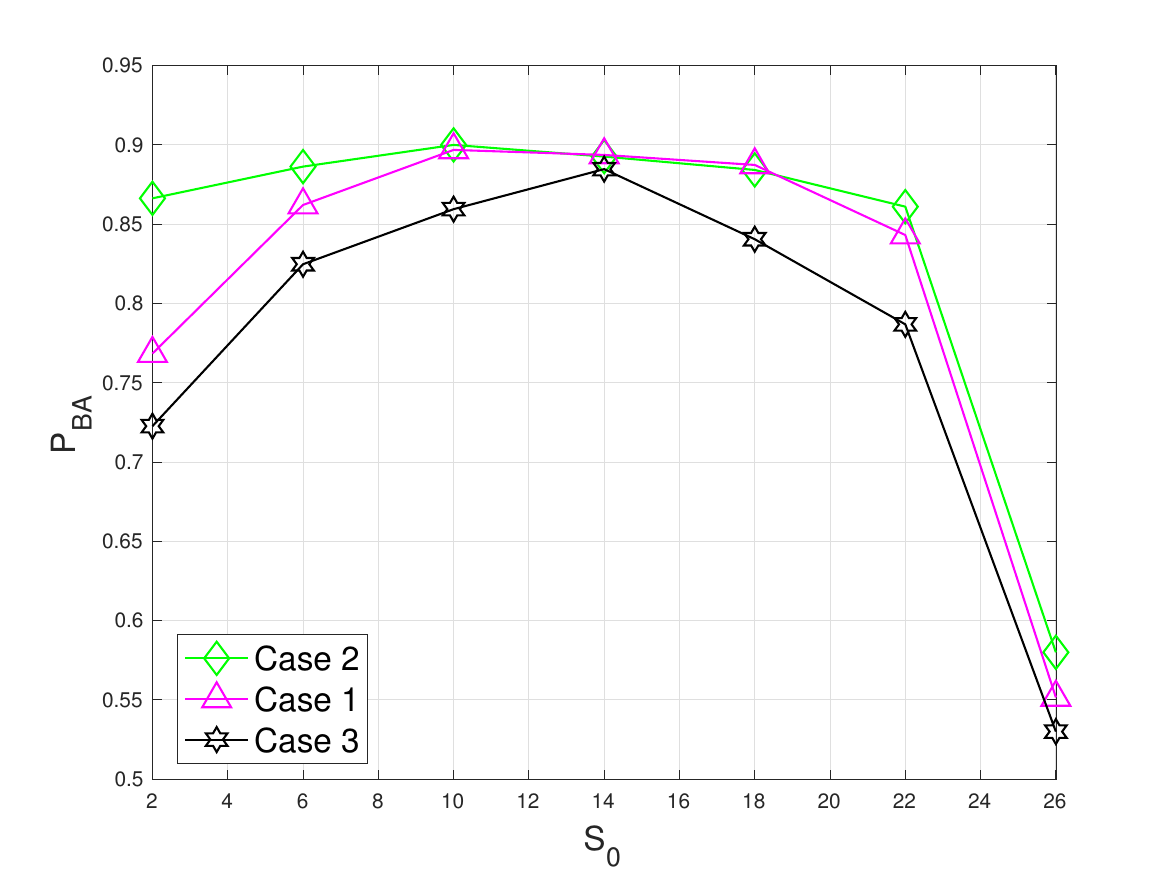}
\caption{$P_\text{BA}$ versus $S_0$ ($\gamma_\Bs=\gamma_\text{J}=1$, 
$Q=100$, and $\text{SJR}=-5$ dB).}
\label{fig:fig_4}
\end{figure}

The performance of the proposed anti-jamming BA scheme as a function of $S_0$
is reported in Fig.~\ref{fig:fig_4}. We remember that $S_1=28-S_0$ in our 
simulation setting.
Results show that there is a significant performance degradation for 
$S_0 < 10$ and $S_0 >18$. The value of $S_0$ impacts on the estimation
accuracy of the jammner-plus-noise power (see Step~2). Values too small
of $S_0$ lead to an unreliable estimate $\widehat{P}^{(j, i)}_{0,\perp}(\widetilde{s})$
of  $P^{(j, i)}_{0,\perp}(\widetilde{s})$ [see eqs.~\eqref{eq:object-00}
and \eqref{eq:object-00-est}] and, thus, involve an inaccurate jamming-plus-noise
cancellation in the proposed NNLS optimization problem \eqref{eq:nnls-perp}.
On the other hand, the value of $S_1$ represents the number of OFDM symbols (per 
each beacon slot and per each subcarrier) collected in Step~3 for building
the estimates $\widehat{P}^{(j, i)}_{1,\perp}(\widetilde{s})$ 
in \eqref{eq:object-11-est} to be used in  \eqref{eq:nnls-perp}.
Values too large of  $S_0$ implies values too small of $S_1$, hence providing poor 
NNLS performance.

\subsection{Probability of successful BA versus number of beacon slots $Q$ and SJR}
\label{sec:3}

\begin{figure}[t!]
\centering
\includegraphics[width=\columnwidth]{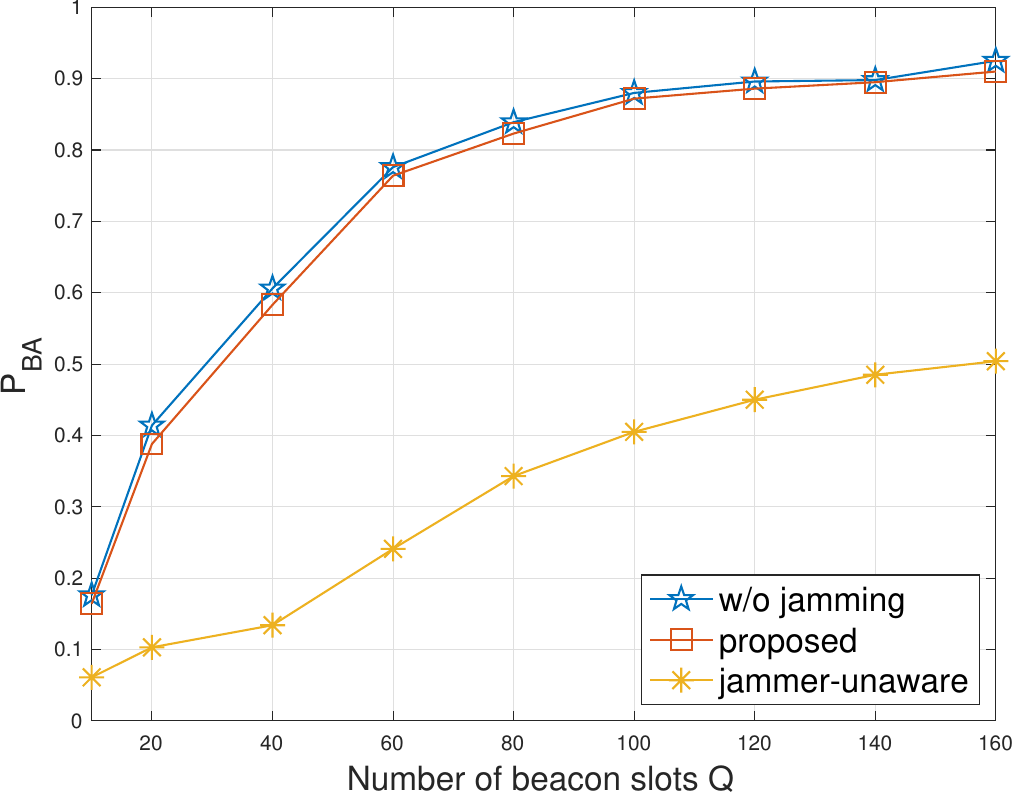}
\caption{$P_\text{BA}$ versus number of beacon slots $Q$ (Case 1,
$\gamma_\Bs=\gamma_\text{J}=1$, and $\text{SJR}=-5$ dB).}
\label{fig:fig_5}
\end{figure}

\begin{figure}[t!]
\centering
\includegraphics[width=\columnwidth]{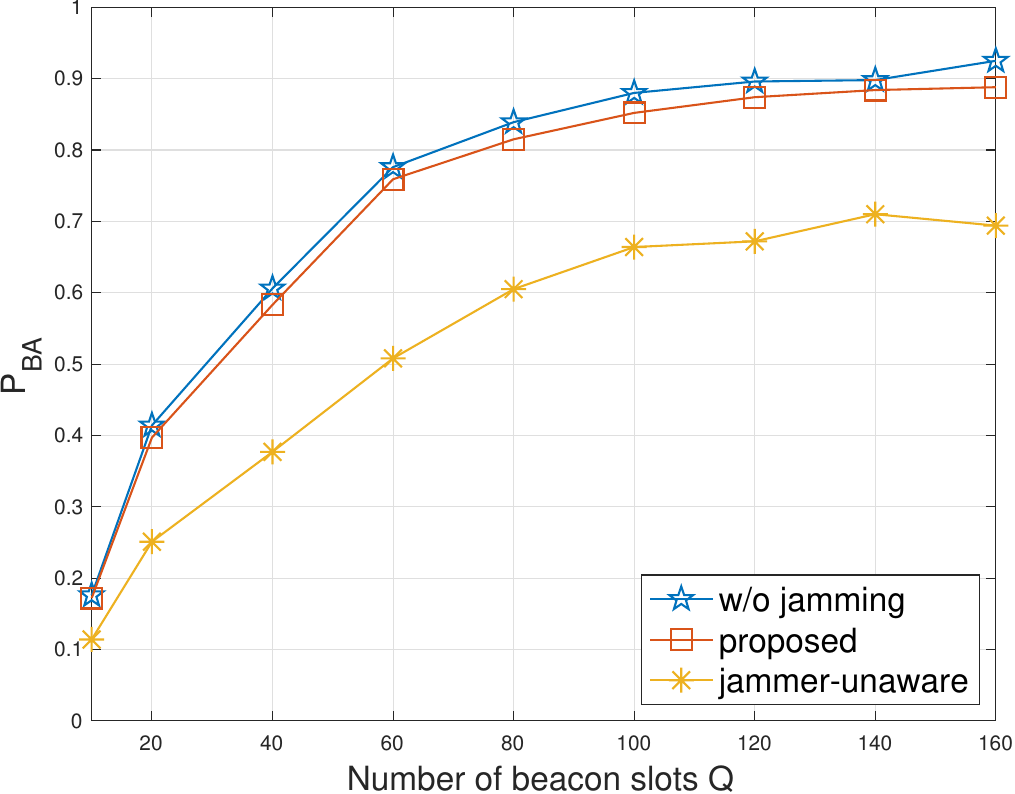}
\caption{$P_\text{BA}$ versus number of beacon slots $Q$ (Case 2, 
$\gamma_\Bs=\gamma_\text{J}=1$, and $\text{SJR}=-5$ dB).}
\label{fig:fig_6}
\end{figure}

\begin{figure}[t!]
\centering
\includegraphics[width=\columnwidth]{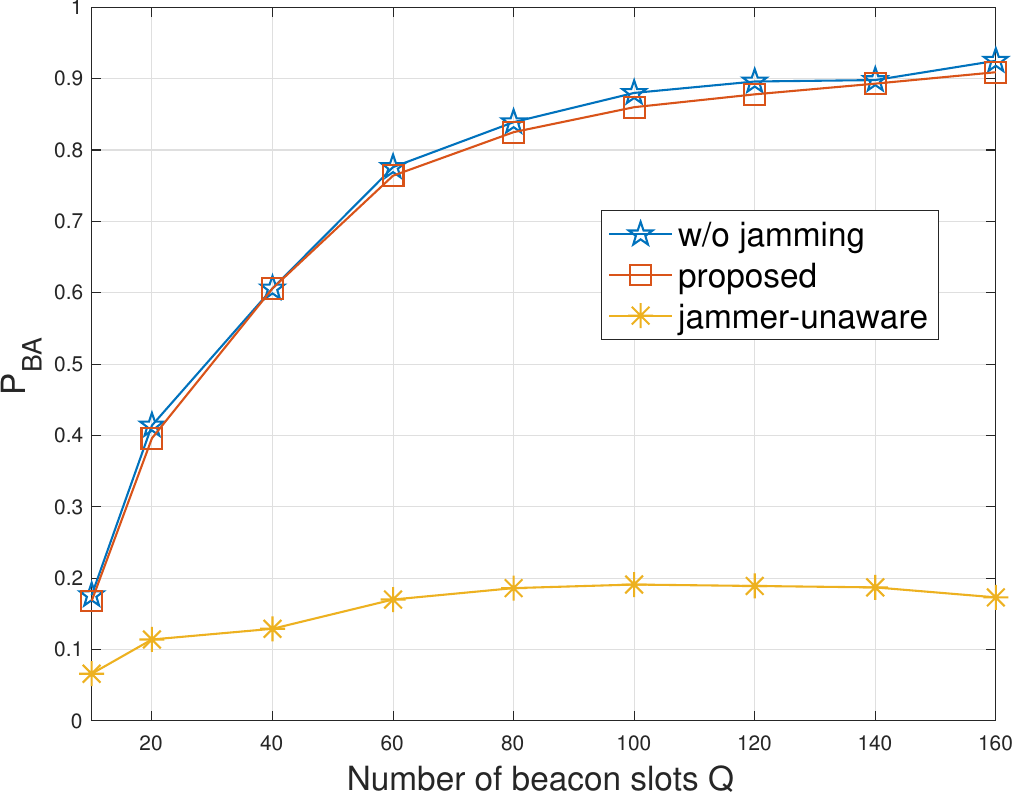}
\caption{$P_\text{BA}$ versus number of beacon slots $Q$ (Case 3,
$\gamma_\Bs=\gamma_\text{J}=1$, and $\text{SJR}=-5$ dB).}
\label{fig:fig_7}
\end{figure}

We report in Figs.~\ref{fig:fig_5}, \ref{fig:fig_6}, and \ref{fig:fig_7}  
the BA performance as a function of the number of beacon slots $Q$.
Additionally, Figs.~\ref{fig:fig_8}, \ref{fig:fig_9}, and \ref{fig:fig_10} depict the probability 
of successful BA as a function of the SJR.
It is seen that, as predicted by our analysis, the performance of the jammer-unaware 
strategy (see Section~\ref{sec:conv}) is very poor when the jamming power 
is equal to or greater than the legitimate signal power,
and successful BA is ensured only when $\text{SJR}> 5$ dB. 
Moreover, the adverse impact of the jamming attack is less  burdensome
in the case of omnidirectional jamming codebook, since 
each beam pattern of the jammer probes simultaneously all the directions, 
thereby spreading the total power in the spatial domain.
Remarkably, the proposed anti-jamming strategy allows to achieve
performance that is very close to that of the ideal case when
there is no jamming attack, thus demonstrating that almost perfect
jammer cancellation is obtained through the proposed three-step
procedure developed in Section~\ref{sec:random-probing}.
Finally, it is apparent that, when the jammer transmits by using the same beamforming codebook
of the BS (Case 3), the jamming-unaware BA
approach is vulnerable to the jamming attack even when the SJR is as high as $5$ dB. On the
other hand, the proposed solution is completely robust with respect to the choice of the jamming
codebook by being able to successfully reject the jamming contribution also in the worst Case 3.


\section{Conclusions and directions for future work}
\label{sec:concl}

We studied the problem of launching a jamming attack during the BA phase between
the BS and users that wish to access the 5G MMW network.
The considered jammer is smart in the sense that it is able to exploit the same 
spatial time-frequency resources that are publicly known to be used by the BS.
In this case, a jamming-unaware approach is not able to ensure successful BA 
between the BS and the legitimate user. 
We proposed a novel BA procedure based on randomized probing and
jammer cancellation, which guarantees performance very close
to that achieved in the absence of a jamming attack.

An interesting research subject consists of considering 
a smart jammer that is able to modify the attack pattern according to
the transmission features of the targeted communication links.
For instance, the jammer might acquire information regarding the
partition of each beacon slot and it may exploits such a knowledge 
to degrade the power estimation process in Step~2. 
In this case, robust solutions have to be developed that allow
to adaptively reconfigure beacon partition and/or     
to use more advanced interference cancellation techniques, 
e.g., independent component analysis.


\appendix

Several conditions on $\G_\Bs$ are known to ensure
that the sparse vector $\xibold_\Bs$ can be estimated from 
the measurement vector $\bm p$. In general, the NNLS problem 
\eqref{eq:nnls} can be ill-posed  if the condition
\be 
\text{$\exists \boldsymbol \alpha \in \Rset^{\Mtilde \Nutilde Q}$ such that $\G_\Bs^\trasp \, \boldsymbol \alpha > \zerobd_{M_\Bs \Nu}$}
\label{eq:cond-1}
\ee
does not hold (see, e.g., \cite{Sla.2013}).
Condition \eqref{eq:cond-1} requires the columns of $\G_\Bs$ 
be contained in the interior of a half-space
containing the origin. Such a condition is fulfilled by 
the transmit beamforming codebook 
\eqref{eq:gtildesimpl}.  

Let $\boldsymbol \beta \in \Rset^{M_\Bs \Nu}$ and $\mathcal{N} \subset \{1,2, \ldots, M_\Bs \Nu\}$
be a subset. We denote with $\boldsymbol \beta_\mathcal{N} \in \Rset^{M_\Bs \Nu}$
the restriction of $\boldsymbol \beta$ to $\mathcal{N}$, i.e., 
$\{\boldsymbol \beta_\mathcal{N}\}_n=\{\boldsymbol \beta\}_n$
for $n \in \mathcal{N}$ and $\{\boldsymbol \beta_\mathcal{N}\}_n=0$ otherwise.
The matrix  $\G_\Bs$ is said \cite[Def.~4.21]{Rau} to satisfy the {\em $\ell_2$-robust 
nullspace property of order
$\kappa_\Bs$} with parameters $\rho \in (0,1)$ and $\varsigma > 0$ if
\be
\|\boldsymbol \beta_\mathcal{N}\|_2 \le \frac{\rho}{\sqrt{\kappa_\Bs}} \, 
\| \boldsymbol \beta_{\overline{\mathcal{N}}}\|_1 + \varsigma \, \|\G_\Bs \, \boldsymbol \beta\|_2 \quad \forall \boldsymbol \beta \in \Rset^{M_\Bs \Nu}
\label{eq:nsp}
\ee
for any subset $\mathcal{N} \subset \{1,2, \ldots, M_\Bs \Nu\}$ with 
$|\mathcal{N} | \le \kappa_\Bs$, 
where $\overline{\mathcal{N}}$ is the complement of 
$\mathcal{N}$  in $\{1,2, \ldots, M_\Bs \Nu\}$.
Property \eqref{eq:nsp} implies that no $\kappa_\Bs$-sparse vectors lie in the
nullspace of $\G_\Bs$. 
It is readily seen from \eqref{eq:gtildesimpl}  and \eqref{eq:Gtx} that 
a (nonzero) vector $\boldsymbol \beta \in \Rset^{M_\Bs \Nu}$ does not belong to the
nullspace of $\G_\Bs$ if and only if 
$\left(\bm{1}_{\mathcal{U}_{\Bs}^{(i)}(\widetilde{s})}\otimes \bm{1}_{\mathcal{V}^{(j)}(\widetilde{s})}\right)^\trasp \boldsymbol \beta \neq 0$ or, equivalently,
\be
\sum_{n \in \text{supp}\left(\bm{1}_{\mathcal{U}_{\Bs}^{(i)}(\widetilde{s})}\otimes 
\bm{1}_{\mathcal{V}^{(j)}(\widetilde{s})}\right)} \{\boldsymbol \beta\}_n \neq 0
\ee
for at least one $i \in \{1,2,\ldots,\Mtilde\}$, $j \in \{1,2,\ldots,\Nutilde\}$, 
and $\widetilde{s} \in \{0,1,\ldots, Q-1\}$. This condition is  fulfilled 
with overwhelming probability for a $\kappa_\Bs$-sparse vector $\boldsymbol \beta$. 
We refer to \cite{Kue} for a rigorous proof in the case of 
$0/1$-Bernoulli matrices.

\begin{figure}[t!]
\centering
\includegraphics[width=\columnwidth]{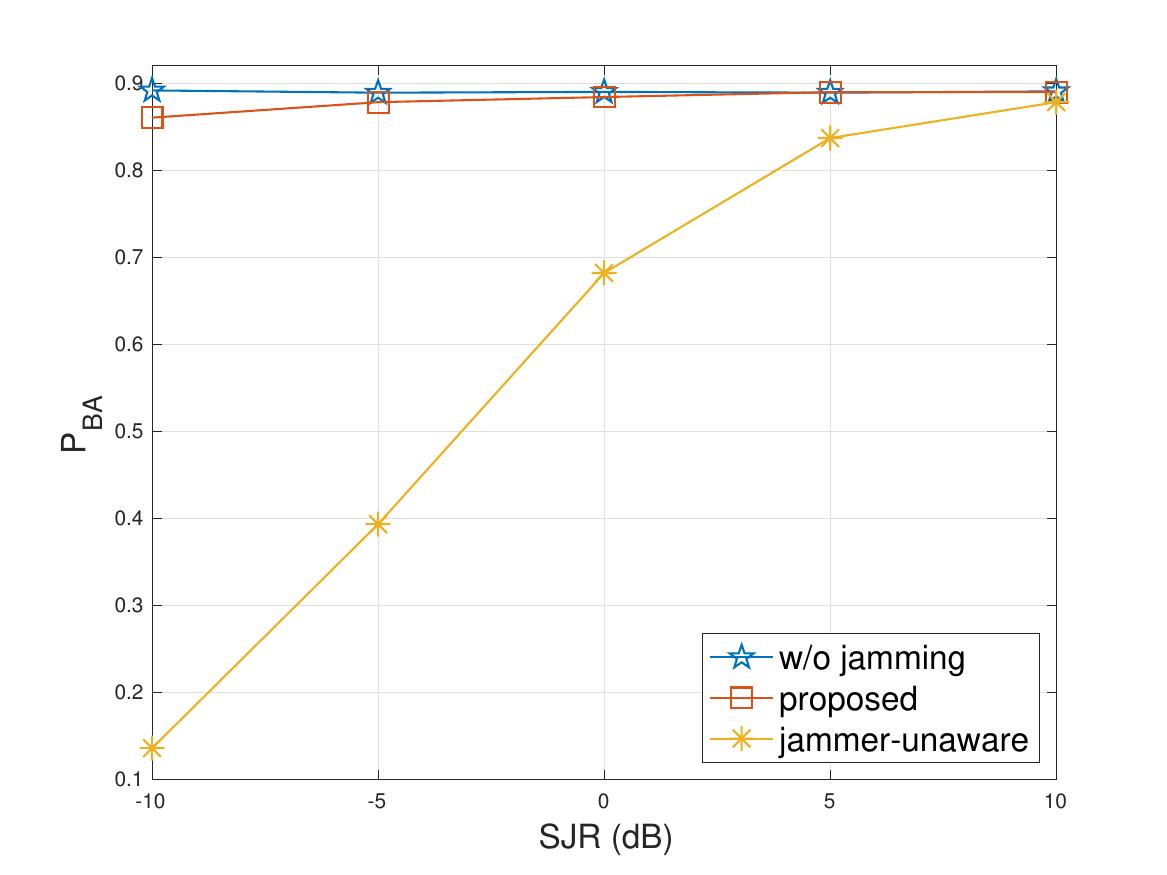}
\caption{$P_\text{BA}$ versus SJR (Case 1,
$\gamma_\Bs=\gamma_\text{J}=1$, and $Q=100$).}
\label{fig:fig_8}
\end{figure}
\begin{figure}[t!]
\includegraphics[width=\columnwidth]{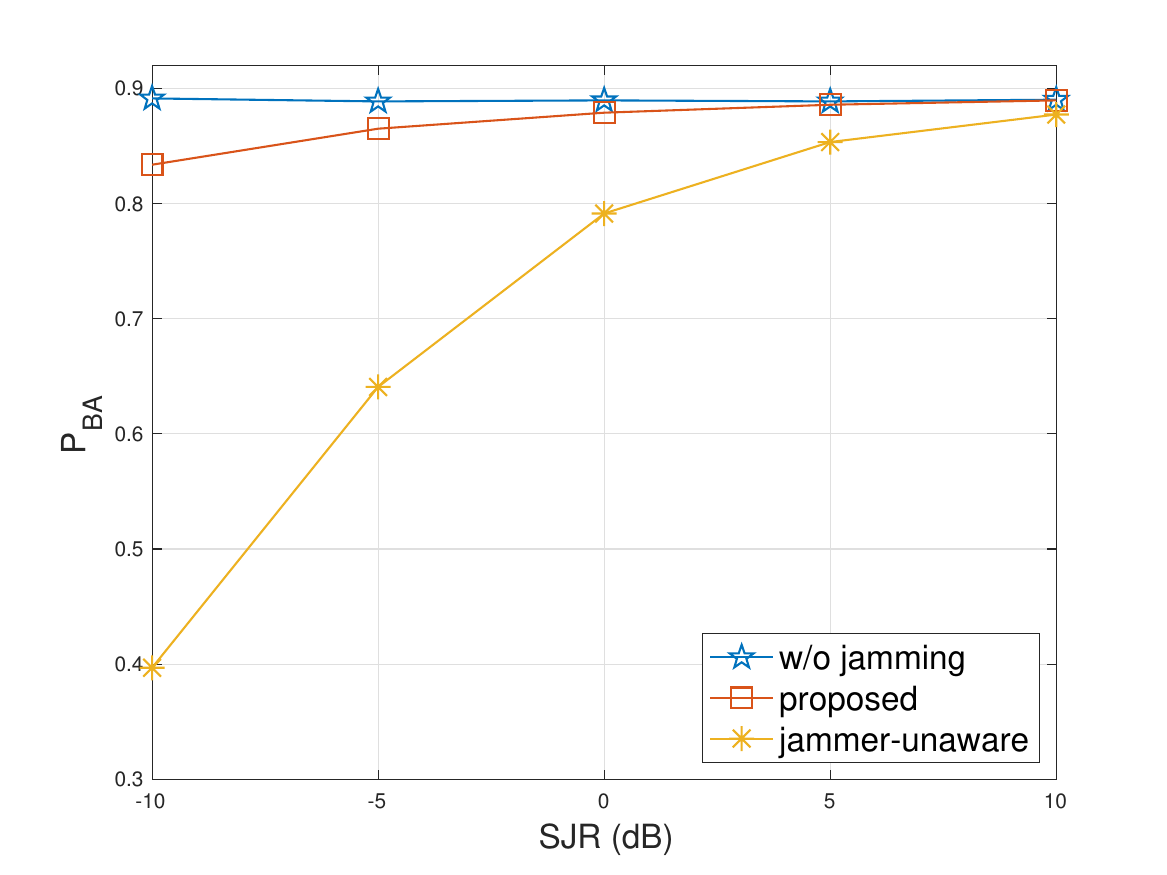}
\caption{$P_\text{BA}$ versus SJR (Case 2, 
$\gamma_\Bs=\gamma_\text{J}=1$, and $Q=100$).}
\label{fig:fig_9}
\end{figure}
\begin{figure}[t!]
\includegraphics[width=\columnwidth]{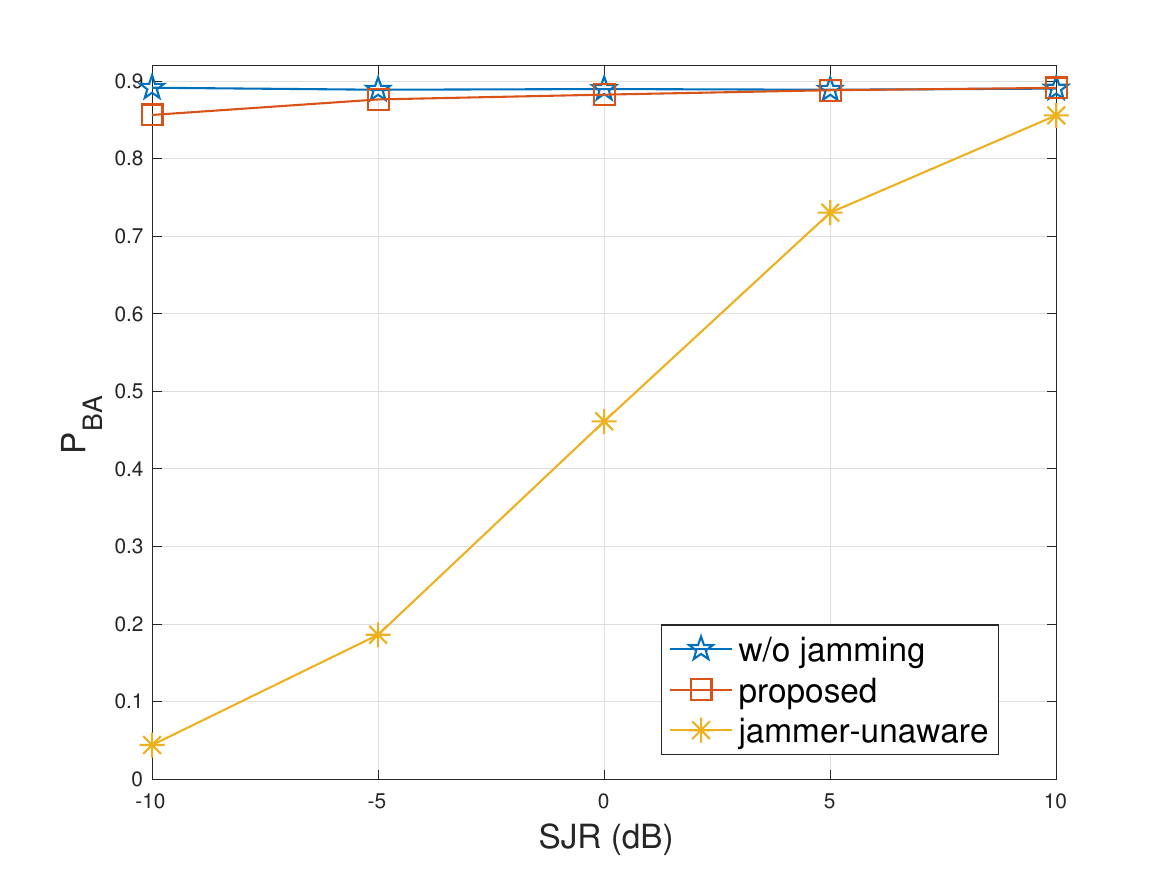}
\caption{$P_\text{BA}$ versus SJR (Case 3,
$\gamma_\Bs=\gamma_\text{J}=1$, and $Q=100$).}
\label{fig:fig_10}
\end{figure}



\begin{IEEEbiography}
[{\includegraphics[width=1in,height=1.25in,clip,keepaspectratio]{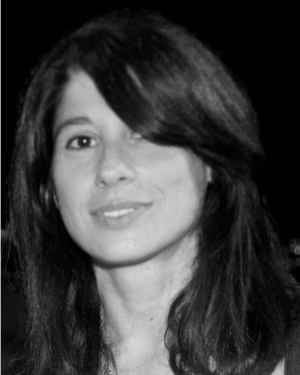}}]
{Donatella Darsena} (M'06-SM'16) received the Dr. Eng. degree summa cum laude in telecommunications engineering in 2001, and the Ph.D. degree in electronic and telecommunications engineering in 2005, both from the University of Napoli Federico II, Italy. From 2001 to 2002, she worked as embedded system designer in the Telecommunications, Peripherals and Automotive Group, STMicroelectronics, Milano, Italy. 
Since March 2005, she has been with the University of Napoli Parthenope, Italy. 
She first served as an Assistant Professor of probability theory 
and, since January 2022, she has served as an Associate Professor of telecommunications with 
the Department of Engineering. 

Her research interests are in the broad area of signal processing for communications, with current emphasis on backscattering communications, space-time techniques for cooperative and cognitive networks, green communications for IoT. 
Dr. Darsena was an Associate Editor for the IEEE COMMUNICATIONS LETTERS from 
December 2016 to July 2019. 
She has served as Associate Editor for IEEE ACCESS since October 2018, 
Senior Area Editor for IEEE COMMUNICATIONS LETTERS 
since August 2019, and Associate Editor for 
IEEE SIGNAL PROCESSING LETTERS since 2020.

\end{IEEEbiography}

\vspace*{-2\baselineskip}

\begin{IEEEbiography}[
{\includegraphics[width=1in,height=1.25in,clip,keepaspectratio]{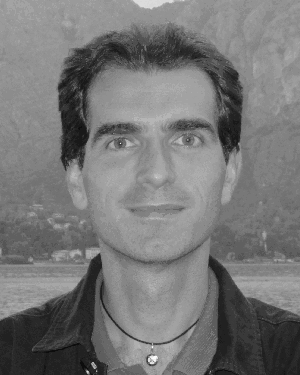}}]
{Francesco Verde}(M'10-SM'14) was born in Santa Maria Capua Vetere,
Italy, on June 12, 1974. He received the Dr. Eng. degree
\textit{summa cum laude} in electronic engineering
from the Second University of Napoli, Italy, in 1998, and the Ph.D.
degree in information engineering
from the University of Napoli Federico II, in 2002.
Since December 2002, he has been with the University of Napoli Federico II, Italy. He first served as an Assistant Professor of signal theory and mobile communications
and, since December 2011, he has served as an Associate Professor of telecommunications with the Department of Electrical Engineering and Information Technology.
His research activities include reflected-power communications, 
orthogonal/non-orthogonal multiple-access techniques, wireless systems optimization, and 
physical-layer security.

Prof. Verde has been involved in several technical program committees of major IEEE conferences in signal processing and wireless communications.
He has served as Associate Editor for IEEE TRANSACTIONS ON COMMUNICATIONS since 2017 and Senior Area Editor of the IEEE SIGNAL PROCESSING LETTERS since 2018. He was an Associate Editor of the IEEE TRANSACTIONS ON SIGNAL PROCESSING (from 2010 to 2014) and  IEEE SIGNAL PROCESSING LETTERS (from 2014 to 2018), as well as Guest Editor of the EURASIP Journal on Advances in Signal Processing in 2010 and SENSORS MDPI in 2018-2022. 
\end{IEEEbiography}

\end{document}